\documentclass[usenatbib]{mnras}

\usepackage{newtxtext,newtxmath}
\usepackage[T1]{fontenc}
\usepackage{ae,aecompl}

\usepackage{graphicx}	% Including figure files
\usepackage{amsmath}	% Advanced maths commands
\usepackage{multicol}
\usepackage{hyperref}
\newcommand{\hi}{H{\sc i}}
\usepackage{ulem, soul}
\title[Effect of errors in observing EoR]{Observing the Reionization : Effect of Calibration and Position Errors on Realistic Observation Conditions}

\author[A Mazumder et al.]{
Aishrila Mazumder,$^{1}$\thanks{E-mail: aishri0208@gmail.com}
Abhirup Datta,$^{1}$
Arnab Chakraborty,$^{1,2,3}$
Suman Majumdar$^{1,4}$
\\
$^{1}$Department of Astronomy, Astrophysics and Space Engineering, Indian Institute of Technology Indore, Indore 453552, India\\
$^{2}$Department of Physics, McGill University, 3600 rue University, Montreal, QC H3A 2T8, Canada\\
$^{3}$McGill Space Institute, McGill University, 3550 rue University, Montreal, QC H3A 2A7, Canada\\
$^{4}$Department of Physics, Blackett Laboratory, Imperial College, London SW7 2AZ, U. K.
}
\date{Accepted XXX. Received YYY; in original form ZZZ}

\pubyear{2015}

\begin{document}
\label{firstpage}
\pagerange{\pageref{firstpage}--\pageref{lastpage}}
\maketitle

\begin{abstract}

Observation of the redshifted 21-cm signal from Cosmic Dawn and Epoch of Reionization is a challenging endeavor in observational cosmology. Presence of orders of magnitude brighter astrophysical foregrounds and various instrumental systematics increases the complexity of these observations. This work presents an end-to-end pipeline dealing with synthetic interferometric data of sensitive radio observations . The mock sky model includes the redshifted 21-cm signal and astrophysical foregrounds. The effects of calibration error and position error in the extraction of the redshifted 21-cm power spectrum has been simulated. The effect of the errors in the image plane detection of the cosmological signal has also been studied. A comparative analysis for array configurations like the SKA1-Low, MWA and HERA has been demonstrated. The calibration error tolerance of the arrays, under some assumptions about the nature of the systematic components, is optimally found to be $\sim 0.01\%$ for the detection of the signal. For position errors, an offset of $\gtrapprox 5\arcsec$ makes the residual foregrounds obscure the target signal. These simulations also imply that in the SKA-1 Low performs marginally better than the others in the image domain, while the same is true for MWA in the power spectrum domain. This is one of the first studies that compares performance of various radio telescopes operating under similar observing conditions towards detecting the cosmological signal. This end-to-end pipeline can also be extended to study effects of chromatic primary beam, radio frequency inferences, foregrounds with spectral features, etc.

\end{abstract}


%%%%%%%%%%%%%%%%%%%%%%%%%%%%%%%%%%%%%%%%%%%%%%%%%%

%%%%%%%%%%%%%%%%% BODY OF PAPER %%%%%%%%%%%%%%%%%%

\section{Introduction}

The thermal history of the evolution of the Universe can be traced back to the surface of the last scattering with the help of the Cosmic Microwave Background (CMB). The evolutionary history of the Universe between the post-recombination era until the period after the neutral intergalactic medium (IGM) became ionized once again remains largely unexplored with observational data. Theoretical models suggest that the Universe's first structures- the earliest stars and galaxies formed during this period (viz. redshifts between $\sim$ 30 and $\sim$ 12). The first stars, generated at Cosmic Dawn (CD), formed due to the small-scale fluctuations in matter density arising from the gravitational instabilities. Ionizing radiations produced from these objects ionized the IGM causing the last phase transition in the evolution of the universe \citep{loeb2001}. This phase transition period is called Epoch of Reionization (EoR) (for comprehensive reviews see \citet{Furlanetto2006, morales2010,Pritchard2012, loeb_and_furlanetto, BARKANA20161,DAYAL20181}). Using quasar absorption spectrum \citep{fan2006} and Thompson scattering optical depth \citep{Planck2018I}, the extended Epoch of Reionization (EoR) is constrained between the redshift interval of $6 \lesssim z \lesssim 15$.

Understanding the conditions prevailing during these early stages of the Universe requires a reliable probe that can trace thermal evolution history. The hyperfine transition line of neutral hydrogen, with a rest wavelength of 21-cm, is the most promising probe into the IGM of these early cosmic epochs \citep{Field1958, Field1959a, Field1959b}. The contrast of the brightness temperature corresponding the 21cm line transition against the CMB gives a differential brightness temperature which is the observable through the all sky-averaged "global signal" (using a single total power radio telescope) or as fluctuations in the spatial correlations (using radio interferometers) \citep{Madau1997}. Radio interferometric observations are necessary for estimation of astrophysical parameters and constraining the nature of reionizing sources. Detection of an absorption trough in the global signal experiment centered at 78 MHz was reported by the Experiment to Detect Global Epoch of Reionization Signature (EDGES) team \citep{Bowman2018}. Besides EDGES, single antenna experiments like BIGHORNS \citep{bighorns}, SCI-HI \citep{sci-hi}, SARAS \citep{saras}, LEDA \citep{leda} are also aiming to detect the global signal but are yet to report any detection.

The most sensitive operational interferometers like the GMRT, MWA, LOFAR, HERA have all set upper limits on the power spectrum (PS) amplitude of the signal \citep{pacgia,mwa1, LOFAR,mwa2,lofar2, 2021arXiv210802263T} but and no confirmed detection of the cosmological \hi\ 21cm signal. The cosmological signal is prone to contamination due to bright astrophysical foregrounds \citep{bharadwaj2005,jelic2008,jelic2010,zahn2011, Chapman2015, samir,arnab2,aishrila2020}. Other sources of contamination are Earth's ionosphere and several instrumental systematics.  Substantial amount of focused research has taken place over the past decade to understand and quantify each components of these contamination, namely  instrument model \citep{delera2017,trott2017,Joseph2018,Li2018}, calibration \citep{offringa2015,barry16,trottwayth2016,ewall2017,Patil2017,dillon18,Kern2019,jais}, foreground avoidance or removal \citep{Datta2010,Trott2012,chapman2014,Vedantham2012,Thyagarajan2015a,Thyagarajan2015b,mertens2018,hothi}, PS estimation \citep{Parsons2010, Parsons2012, liu2014, tge14,tge16, Offringa2019, barry2019} and Earth's ionosphere \citep{jordan2017,Trott2018}. Using these advancements, highly sensitive next generation interferometers like the Hydrogen Epoch of Reionization Array (HERA, \citealt{DeBoer2017}) and the Square Kilometer Array (SKA1-Low, \citealt{Koopmans2015}) are expected to detect the 21-cm signal and characterize the multi-redshift PS, leading to tighter constrains on the astrophysical parameters in the early Universe.  The upcoming SKA1-Low is being designed to be sensitive enough to detect the PS precisely. Additionally, it is also projected to be able to create tomographic images of the HII regions \citep{Mellema2014}. 
The EoR signal is inherently isotropic in spatial wavenumber (k) space. Additionally, it also shows spectral structure. Thus, despite the spectrally smooth foreground contamination can be distinguished from the foreground contaminants \citep{Morales2004, Morales2006, Bowman2009,liu2011, Parsons2012, Pober2013,dillon2013}. The spectrally smooth nature of the foregrounds, along with the inherent instrumental chromaticity, keeps the contamination confined to the "wedge" in the cylindrical Fourier space (i.e., in the 2D PS). The region outside the wedge where the foregrounds are subdominant compared to the signal is called the "EoR window" \citet{Morales2012}. However, the interaction of astrophysical foregrounds and the instrument causes the wedge power to leak into the clean modes of the window- an effect called "mode mixing" \citep{Morales2012}. If not properly mitigated, mode mixing is likely to confuse the detection.

 One of the major limiting systematic that causes the problem in cosmological signal detection is improper calibration. In general, the most common calibration approach for radio astronomy is "sky-based" calibration. Since Cosmic Dawn/Epoch of Reionization (EoR) observations are limited to shorter baselines, these observations suffer from poor angular resolutions and higher confusion noise. This may lead to inaccurate sky models for calibration. Errors or inaccuracies in this process propagate as residual errors and ultimately hinder the target cosmological signal detection \citep{Datta2009, Datta2010,barry16, ewall2017}.
The other calibration approach being explored by various interferometers is the redundant calibration approach . Telescopes  like the HERA use the redundant calibration approach. In this approach, repeated simultaneous measurements of the interferometers' redundant baselines can simultaneously solve for the incoming sky signal and instrumental parameters. The redundant method calibrates for real visibility by considering the prior that sky visibilities are equal for redundant baselines, having the advantage of not requiring a sky model for performing gain solutions for antennas. Nevertheless, this works only for highly regular arrays and identical antenna beam patterns \citep{Byrne2019, Byrne2020}.

It must be pointed out that irrespective of the accuracy of the calibration approach employed, the ultimate limitation is set by the signal-to-noise ratio (SNR). It has been shown in \citet{Datta2009}, that calibration errors $\sim$0.1\% causes a drastic lowering of the dynamic range, thereby obscuring the signal\footnote{Dynamic range is the ratio of the peak flux in the image and the RMS noise in a source-free region. \citet{Datta2009} showed that with calibration error of $\sim$0.1\%, the dynamic range lowers to 10$^5$ compared to the required 10$^8$}. Thus, studies are necessary for estimating the extent of the tolerance of the imperfections, which potentially contribute to the noise that obscure the detection of the faint cosmological signal. This paper aims to quantify this level of tolerance (or the accuracy of the calibration algorithm) required for successful detection of the redshifted 21-cm signal from the cosmic reionization using various sensitive telescopes.

Astrophysical foregrounds like diffuse galactic synchrotron emission, extra-galactic free-free emission, and compact sources substantially affect the interferometric data sets used for EoR signal recovery. Currently, most studies focus on determining the best strategy- for avoidance, suppression, or subtraction for handling the foregrounds. While diffuse emission is a significant contaminant at large angular scales, EoR data sets also require the careful handling of the extragalactic point sources.  Inaccurate removal of bright, compact sources can lead to large residuals in the data and obscures the detection of the cosmological signal. Besides ionosphere, incomplete UV-plane coverage gives rise to large sidelobes that contaminate the data. All these corruption effects lead to constructing an imperfect sky model, which causes residuals that adversely affects the cosmological signal recovery. The study on the impact of incomplete sky model has been done by \citet{Datta2009, Datta2010, Trott2012, barry16, Byrne2019}. However, there is still a lack of detailed study on how the effects compare for different telescope arrays targeting the cosmological 21-cm signal from reionization.

Our current work presents a proof of concept on developing an end-to-end pipeline to simulate radio interferometric observations using a realistic sky model and telescope observing parameters. The studies have been carried out taking four array configurations - SKA1-Low core (stations around $\sim$ 2000 m of the central station), HERA 350 dish configuration, MWA Phase-1 (MWA1) 128 tile configuration, and MWA 256 tile configuration (MWA-256). 
This paper is organized in the following manner: in Section \ref{Simulation}, the detailed simulation methodology is described. Section \ref{errors} discusses briefly the observational effects considered in this work; Section \ref{formalism} discusses the analyses formalism which is followed by the Section \ref{results}; finally, the paper is concluded Section \ref{conclusion}.

\section{Synthetic Observations}
\label{Simulation}

\begin{figure}
    \centering
    \includegraphics [width=\columnwidth] {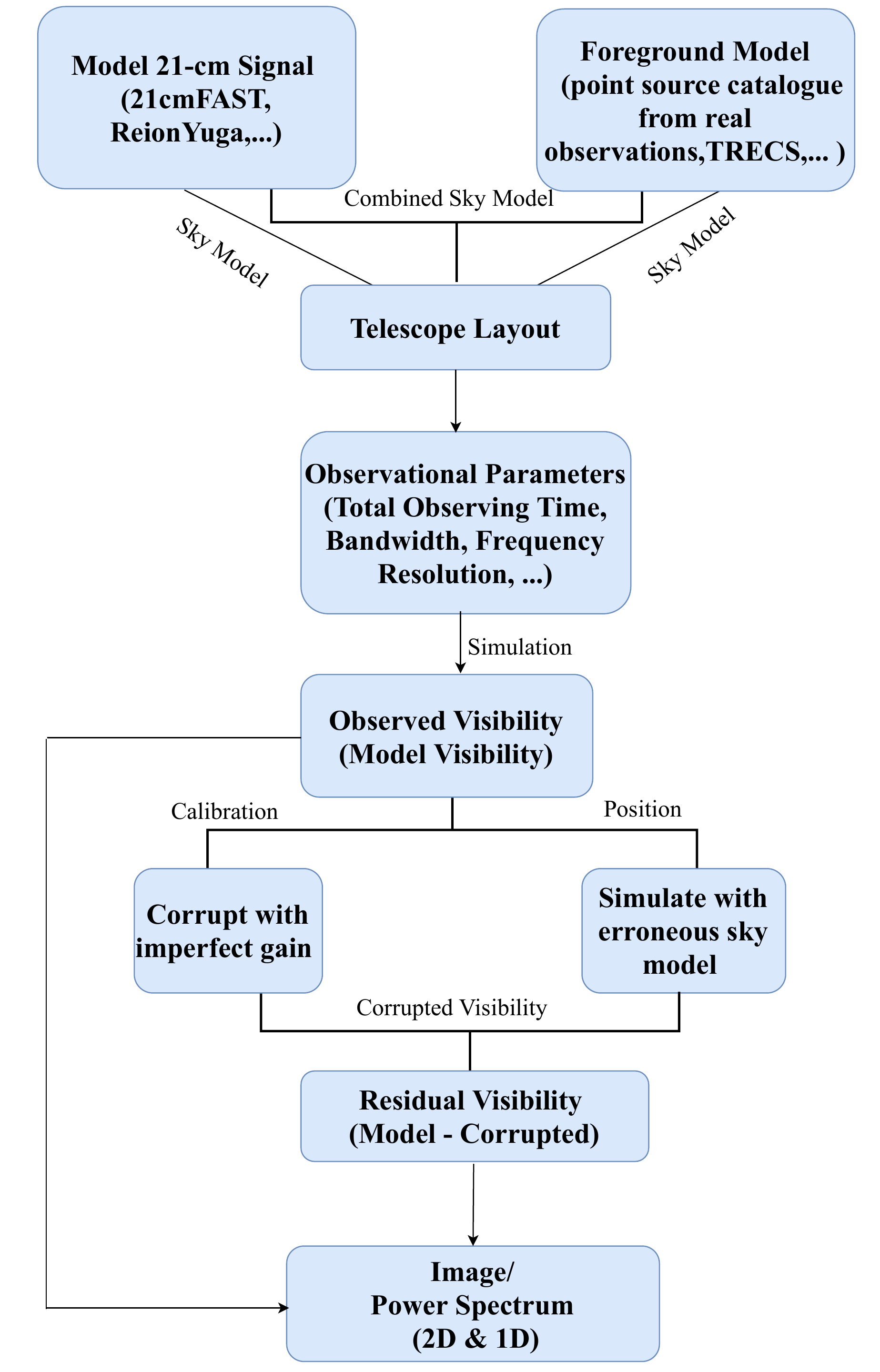}
    \caption{Schematic diagram of the observation pipeline.}
    \label{flowchart}
\end{figure}

This section describes the steps involved in the synthetic observation and analysis pipeline developed. Figure \ref{flowchart} shows the blocks that are used for generating the observations. The various simulation parameters and their respective values are described in Table \ref{paramters}. In the following subsection, each of the parameters in the input block is briefly described.

\begin{table} %[h!tba]
\begin{center}
\caption{Parameter values for the simulations}
\label{paramters}
\begin{tabular}[\columnwidth]{ll}
\hline
\hline
Parameter & Value \\
\hline
Central Frequency & 142 MHz (z$\sim$9) \\
\hline 
Bandwidth & 8 MHz \\
\hline
Number of frequency channels & 64 \\
\hline
Field of view & 4$^\circ$ \\
\hline
Number of array elements: & \\
SKA1-Low & 296\\
HERA & 350\\
MWA Phase-1 & 128 \\
MWA-256 & 256 \\
\hline
Maximum baseline (m): & \\
SKA1-Low & $\sim$2000\\
HERA & $\sim$880\\
MWA Phase-1 & $\sim$2800 \\
MWA-256 & $\sim$5300 \\
\hline
Synthesised beam (arcmin): & \\
SKA1-Low & $\sim$2.5\\
HERA & $\sim$8.1\\
MWA Phase-1 & $\sim$2.6\\
MWA-256 & $\sim$1.3\\
\hline
Thermal Noise (mJy $\mathrm{beam^{-1}}$):\\ 
SKA1-Low & $\sim$0.07\\
HERA & $\sim$0.52 \\
MWA Phase-1 & $\sim$17.59 \\
MWA-256 & $\sim$8.79 \\
\hline
\hline
\end{tabular}
\end{center}
\end{table}

The simulations have been performed using the OSKAR software \citep{oskar}\footnote{\url{https://github.com/OxfordSKA/OSKAR/releases}} package for SKA1-Low configuration and Common Astronomy Software Application (CASA, \citealt{casa})\footnote{\url{https://casaguides.nrao.edu/index.php?title=Main_Page}} for HERA, MWA-1 \& 2. The observations track the sky with phase center at $\alpha$=15h00m00s and $\delta$=-30$^\circ$00\arcmin00\arcsec for 4 hours ($\pm$ 2 HA). The observing bandwidth is 8 MHz, with a channel separation of 125 kHz. For simplicity, the simulations are noise-free. It should also be mentioned that throughout the paper, best fitted cosmological parameters from the Planck 2018 results \citep{Planck2018I} are used: $\Omega_{\mathrm{M}}$ = 0.31, $\Omega_{\Lambda}$ = 0.68, $\sigma_{8}$ = 0.811, $H_{0}$ = 67.36 km $\mathrm{s}^{-1}$ $\mathrm{Mpc}^{-1}$.

\subsection{Telescope Model}
\begin{figure}

    \includegraphics[width=\columnwidth,height=6.5cm]{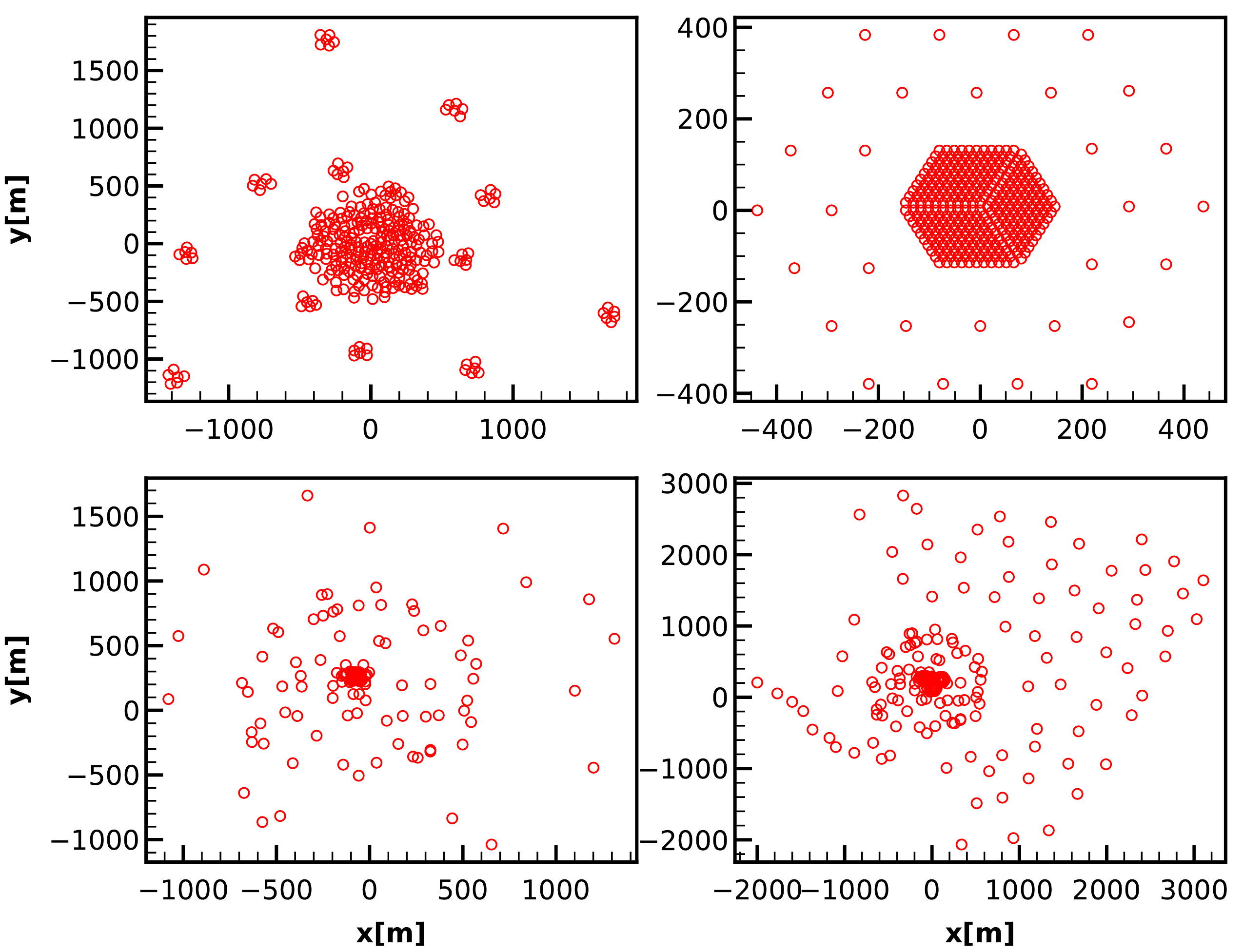}
    \caption{Telescope layouts used in the simulation. Clockwise from top left are SKA1-Low (Low), HERA, MWA-256 \& MWA-I respectively.}
    \label{arrays}
\end{figure}

Though several interferometers are targeting the detection of the redshifted 21-cm signal, this work undertakes a comparative analysis under identical error conditions for four different arrays- SKA1-Low, HERA, and MWA1 \& MWA2. The telescope layouts are shown in Figure \ref{arrays}.

The SKA\footnote{\url{https://www.skatelescope.org/}}, one of the most sensitive upcoming telescopes, is expected to detect the signal statistically \citep{Koopmans2015} and also have just enough signal-to-noise for performing tomography \citep{Mellema2014}. SKA will be built at two sites as two separate arrays, a low-frequency array in Australia (frequency coverage between 50-350 MHz) and a mid-frequency array in South Africa (frequency coverage between 350 MHz to 14 GHz). The redshifted 21-cm signal from the EoR is present at frequencies of up to a few hundred MHz ($\sim$200 MHz at z $\sim$6) and thus will be targeted by the former array (SKA1-Low). While the available documentation states that the SKA1-Low will have 512 "stations" of $\sim$ 40m diameter spread around the core or central station (having a maximum baseline length of $\sim$65 km, \citealt{SKAO_construction}), this work does not explore the entire layout. Since compact baselines will be sensitive to the target signal, all stations within 2 km of the core (i.e., a maximum baseline of 2000m around the central station) are used \citep{ska_design}. This results in 296 stations around the central station, with a synthesized beam size of $\sim$2.5\arcmin at 142 MHz.

HERA\footnote{\url{https://reionization.org/}} is an interferometer designed to search for the fluctuations in the 21-cm signal during the cosmic dawn and reionization \citep{DeBoer2017}. Located in the South African Karoo Radio Astronomy Reserve, the complete array would comprise 350 parabolic dishes of 14m diameter each, spread in a highly redundant configuration (320 in a dense core and 30 outriggers). The instrument is planned to be operational between 100-200 MHz. This work uses the entire 350 tile configuration resulting in a maximum baseline of $\sim$880 m. At 142 MHz, this results in an $\sim$8\arcmin synthesized beam.

MWA\footnote{\url{https://www.mwatelescope.org/}}, an SKA precursor facility, is located at the Murchison Radio Observatory in Australia (adjacent to the SKA1-Low site). The array has an operating frequency range of 70-300 MHz. In the initial deployment stage, the array had 128 square-shaped "tiles" (a $4\times4$ array of dipoles). In Phase 2, the 128 tile configuration was extended to include additional 128 tiles \citep{mwanew}. The upgraded array has a compact configuration with 56 tiles from earlier and 72 new tiles arranged in two compact hexagons, which enhance EoR PS capacity \citep{mwanew}. For this work, two MWA configurations are used- entire Phase 1 with 128 tiles and entire Phase 2 with 256 tiles \footnote{The MWA coordinates are available at \url{https://www.mwatelescope.org/telescope/configurations/phase-ii}}. The synthesized beam-width at 142 MHz are $\sim$2.5\arcmin $\sim$1.3\arcmin respectively for Phases 1 \& 2 respectively.

\subsection{The \hi\ 21-cm Maps}
\label{maps}

We use two semi-numerical approaches, 21cmFAST \citep{21cmFAST, 21cmFASTv3} and ReionYuga (formerly known as Sem-Num) \citep{mondal17,suman2014, tirth09} to generate the redshifted 21-cm signal maps for our simulated observations. Both of these methods are developed independently, and they use different approaches of excursion set formalism to simulate the 21-cm maps from the EoR. Both of these methods have been tested against full radiative transfer simulations and found to be capable of simulating the signal power spectrum at large length scales with a reasonable accuracy \citep{suman2014}. They, however, show significant differences at small length scale power spectrum. Our motivation behind using two separate signal simulations is to test the robustness of the outcome of our data analysis pipeline. It is going to use both extracted image and power spectrum statistics for its evaluation.

21cmFAST generates the 21-cm maps from a matter density field obtained via Zeldovich approximation. It uses a guided excursion set formalism to convert the matter density field at a given redshift into an ionization field which is finally translated into  21-cm brightness temperature fluctuations. The excursion set approach allows it to generate multiple realizations of the 21-cm maps at a very low computing cost compared to a full radiative transfer approach. 

The ReionYuga or Sem-Num (a serial version of ReionYuga), on the other hand, uses a \textit{N}-body simulation to generate the matter distribution at a given redshift and further uses a Friends-of-Friends algorithm to identify collapsed halos in this matter distribution. These halos are then considered the hosts of the first sources of lights (e.g., galaxies, quasars, etc.), which produces the ionizing photons responsible for reionizing the Universe. The density of the neutral hydrogen field (assuming that the hydrogen follows the dark matter distribution) and the ionizing photon field are then estimated at a much coarser grid compared to the original N-body simulation resolution, and using a guided excursion set formalism, an ionization field is created. This ionization field is then converted into a 21-cm brightness temperature field.

For our analysis in this paper, we use the output of an already existing run of Sem-Num, precisely the fiducial light-cone map of \citet{majumdar16}. This light-cone volume has sky plane comoving extent of $500 \times 500\, (h^{-1} {\rm Mpc})^2$ and the line-of-sight or redshift extent of $7.0 \leq z \leq 12.0$. The 21-cm map in this light-cone volume is saved on a grid of size $232 \times 232 \times 562$. We extract a cuboid of grid dimension $232 \times 232 \times 64$ from this light-cone volume and project it to a World Coordinate System (WCS) of the same dimension. This extracted cuboid is used as the signal model for our analysis. Similarly, we generate a 21-cm light-cone cuboid using 21cmFast of the same volume and treat it as our other signal model.  

\begin{figure}
    \includegraphics[width=\columnwidth,height=6cm]{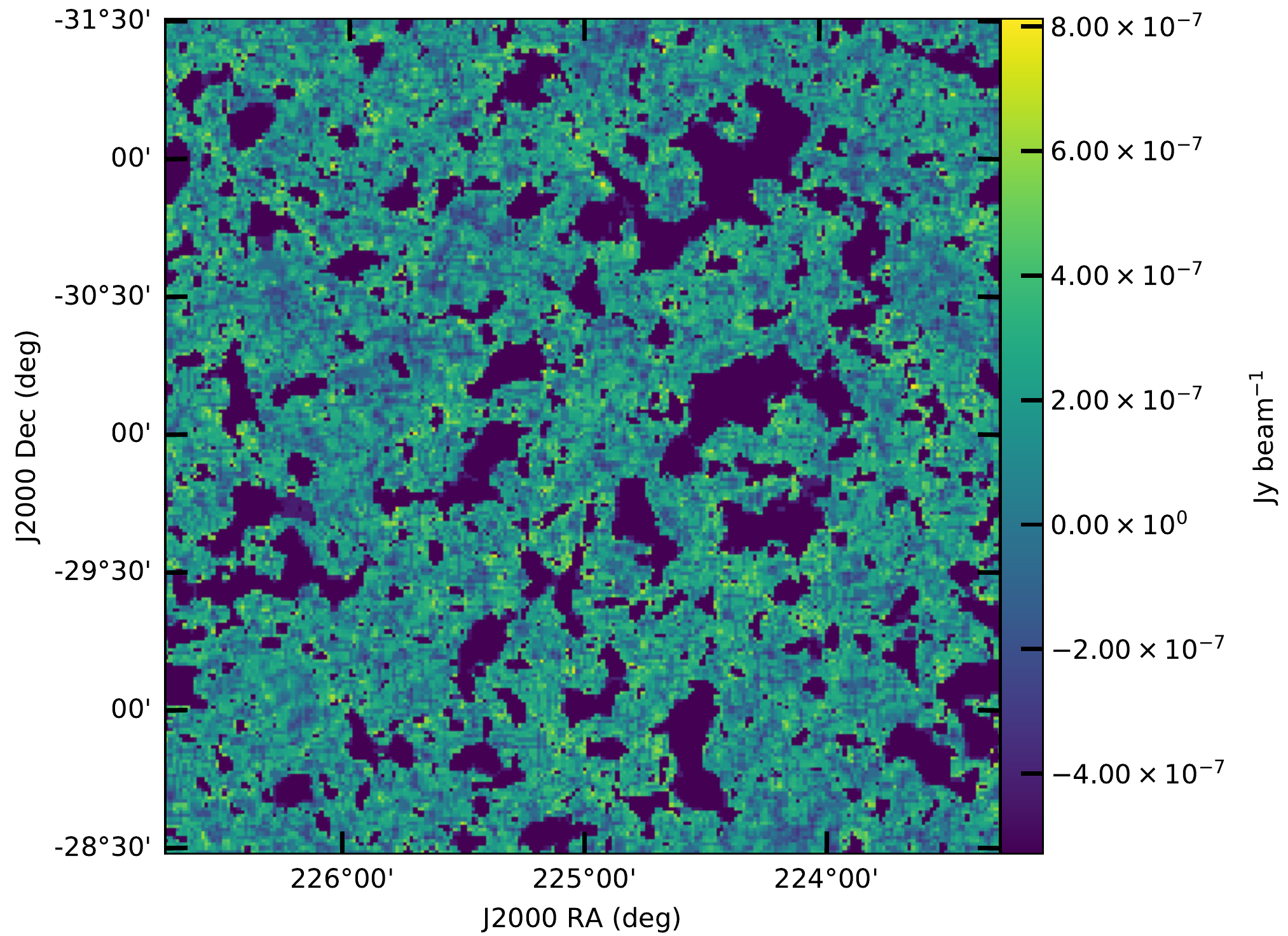}
    \includegraphics[width=\columnwidth,height=6cm]{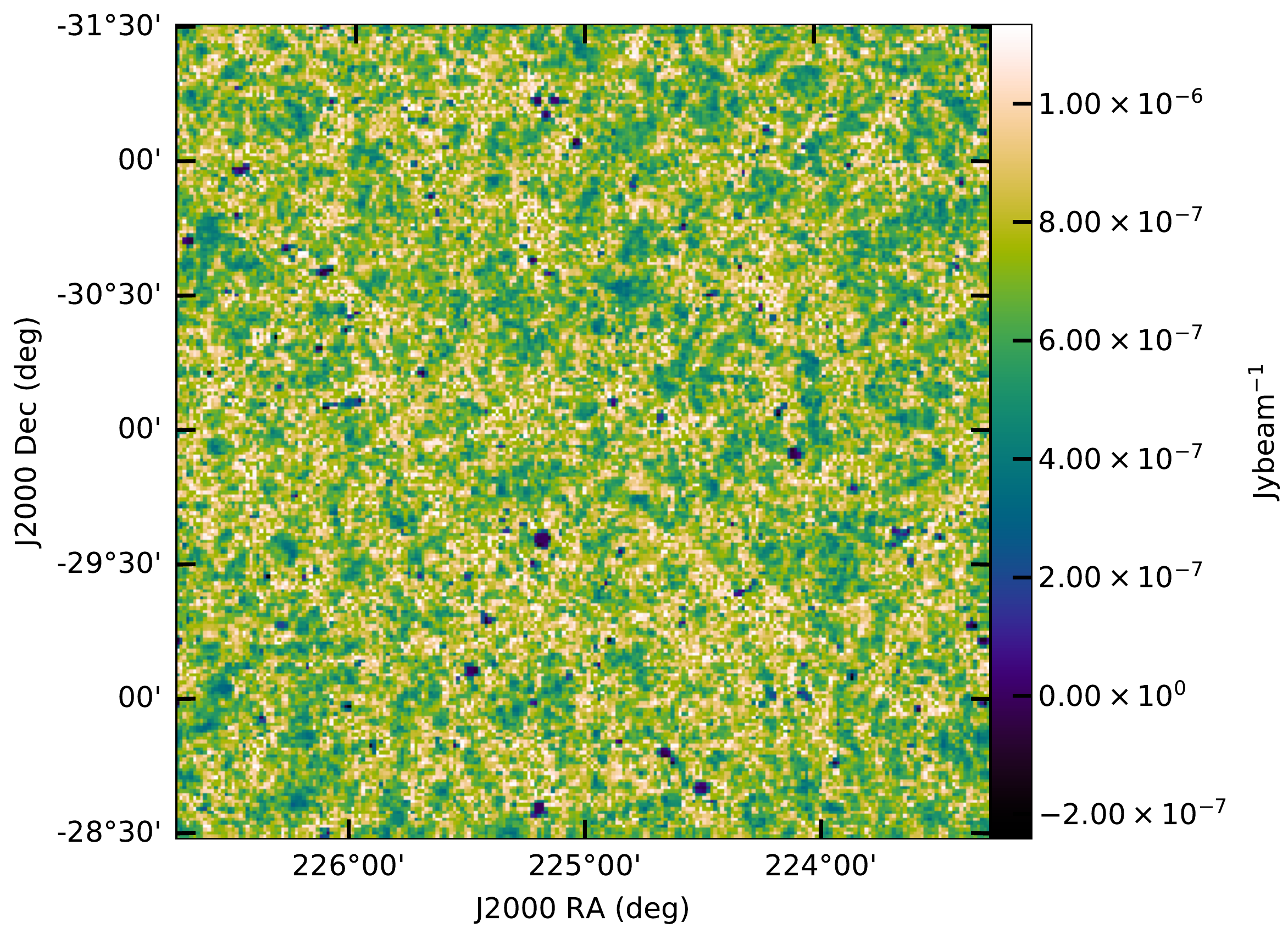}
     \caption{Slice through the \hi\ 21-cm light cone at 142MHz, projected on a WCS using 21cmFAST (top) and SemNum (bottom). The light cones produced have different reionization histories and are thus not identical.}
    \label{hi}
\end{figure}

\begin{figure}
    \includegraphics[width=\columnwidth,height=6cm]{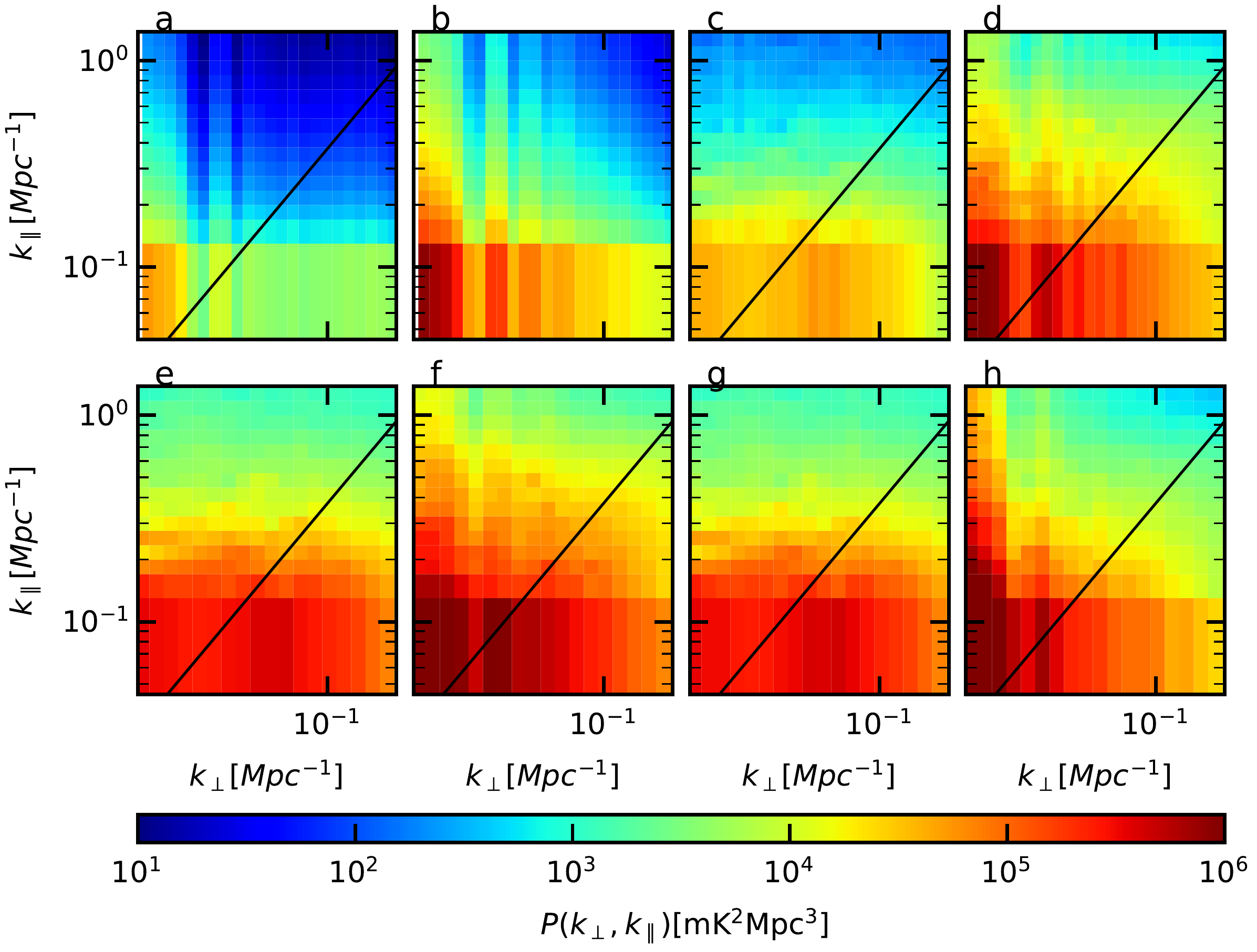}
    \includegraphics[width=\columnwidth,height=7cm]{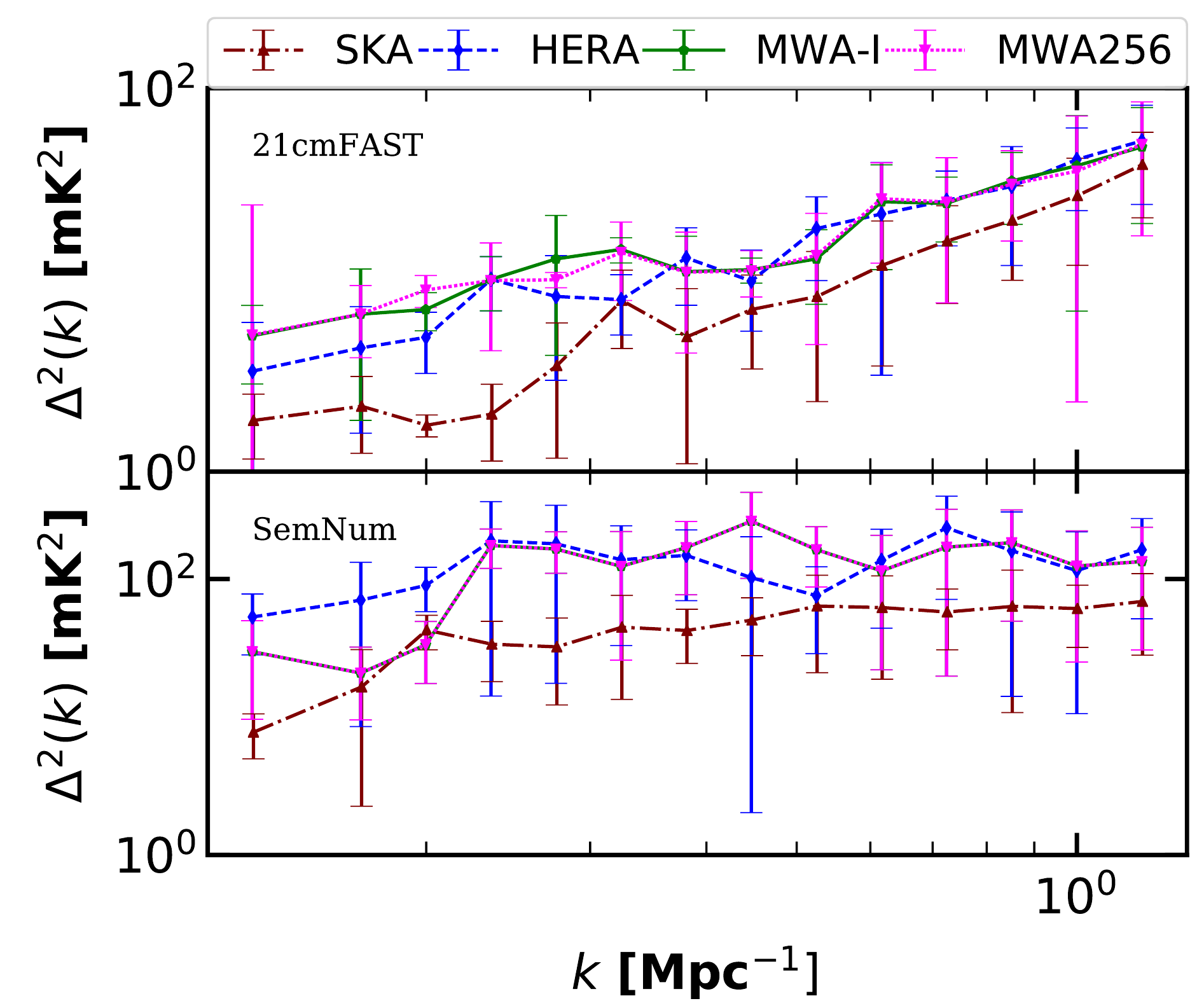}
 
    \caption{\textbf{top}Power spectrum obtained for the two signal models used when observed with the four telescope configurations. The top panel shows the cylindrical PS with SKA1-Low (panel a-21cmFAST, b-Sem-Num), HERA (panel c-21cmFAST, d-Sem-Num), MWA-I (panel e-21cmFAST, f-Sem-Num), MWA-256 (panel g-21cmFAST, h-Sem-Num). \textbf{bottom} The spherical PS with 21cmFAST \textbf{(upper subplot)} and SemNum \textbf{(lower subplot)}. For both subplots, the layouts are SKA (maroon dot-dash line with triangle markers), HERA (blue dashed line with diamond markers), MWA-I (green solid line with pentagon markers) and MWA-256 (magenta dotted lines with inverted triangle markers).}
    \label{semnumps}
\end{figure}

Figure \ref{hi} shows a slice of the signal at the central frequency (142 MHz, z$\sim$9) for 21cmFAST (top panel) and Sem-Num (bottom panel). It is evident from this figure that the reionization histories followed by these two models are not the same; thus, at the same redshift, they arrive at a significantly different level of IGM ionization. Therefore, the levels of resulting 21-cm fluctuations are also considerably different. Figure \ref{semnumps} shows the power spectrum from these two signal models as observed by four different telescope configurations discussed earlier. Here we assume that the observational data contains the signal alone. We follow that the resulting power spectra for the two signal models are not drastically different at the length scales of our interest. As 21cmFAST is a faster simulation to implement, the subsequent analysis has been done using this signal model alone.

\subsection{Foreground Models}
\label{sky}

The foreground used for this work consists of compact sources only. Once again, two separate catalogs have been used - the 400 MHz catalog obtained using uGMRT of the field ELAIS N1 \citep{arnab2} and the Tiered Radio Extragalactic Continuum Simulation (T-RECS, \citet{trecs}) to demonstrate the ability of the pipeline. The sources were redistributed around a declination of -30$^\circ$. The fluxes were converted to 142 MHz values using $S_\nu \propto \nu^{-\alpha}$, with $\alpha$ value of -0.8. The minimum flux value present at 400 MHz was 100 $\mu$Jy, and the highest flux is 0.6 Jy.  T-RECS is a simulation modeling the continuum radio sky in the range 150MHz to 20GHz. It models Active Galactic Nuclei (AGNs) and Star-Forming Galaxies (SFGs), comprising most radio galaxy populations. These simulations consider real data from recent observations to model the sources. This gives realistic cosmological evolution of the luminosity functions, the total intensity number counts, polarized intensity, and clustering properties. The source catalogs are available publicly for 1 deg$^2$, 25 deg$^2$, and 100 deg$^2$. For this work, from the 25 deg$^2$ catalog, sources within $\sim$4\ deg$^2$ having flux cut-off the same as EN1 catalog is chosen (the brightest source in the constructed catalogue has a flux of 1.7 Jy at 150 MHz). Fluxes at 150 MHz were scaled to 142 MHz using $\alpha \approx$ -0.8. The catalog comprises 2522 sources in the chosen field of view  (FoV) and flux cut-off.  

It is to be noted that no diffuse emission model was used in the foreground model to keep the present analysis simple. Future works will incorporate models of the diffuse emission for investing their effect.

\section{Non-ideal observing conditions}
\label{errors}
Observational effects on the cosmological 21-cm signal are numerous. However, this work focuses on calibration errors and position errors. 

Radio interferometers observe the Fourier transform of the sky intensity distribution. The observed visibilities (i.e., spatial coherence function of sources) are calculated by cross-correlating the detected electric field between antenna pairs using digital correlators. The individual antenna (or tile) measures the incident electric field modified by a complex gain factor (arising due to the electronics). Calibration aims to solve for these complex gains. For observations that target the cosmological 21-cm signal, either a "traditional" sky-based technique or a "redundant" calibration technique is used. In the former method, a sky model is used to calibrate instruments \citep{Taylor1999}. The latter is useful for arrays that have a large number of redundant baselines \citep{liu2010}. However, irrespective of the technique used, the basic equation is the measurement equation relating the measured sky visibility to true sky visibility. The simplest form of the equation is :
\begin{equation}
    V{_{ij}^m} = g_{i}(t)g{_j^*}(t)V{_{ij}^{t}}
    \label{me}
\end{equation}
where $V{_{ij}^m}$ is the measured visibility from the antenna pairs i and j, $V{_{ij}^{t}}$ is the true sky visibility and $g_i(t)$ and $g{_j^*}(t)$ are the complex gains of the respective antennae. While gain calibrations aim to solve for these gains, there is a limit to the accuracy of these. The uncalibrated gains (or residual gains) that remain are propagated in the subsequent steps and leads to errors in extracting the target signal. The complex gain can be modeled by 
\begin{equation}
    g{_i} = (a_{i}+\delta{a_i})exp(-i({\phi_i}+\delta{\phi_i}))
    \label{gaineq1}
\end{equation}
where the phase, $\phi_i$ is in radian, and the amplitude, $a_i$ is dimensionless. The terms $\delta{a_{i}}$ and $\delta{\phi_i}$ are the errors in the gain solutions. For the ideal case, $a_{i}$ is 1 and ${\phi_i}$ is 0, giving gain of 1. However, due to the presence of residual gain errors, the error terms will give the resultant gain as 

\begin{equation}
    g{_i} = (1+\delta{a_i})exp(-\delta{\phi_i})
    \label{gaineq}
\end{equation}
An "efficient" calibration algorithm will be one that can minimize $\delta{a_{i}}$ and $\delta{\phi_i}$. The target signal itself is feeble, thus attaining an SNR $>$1 becomes a challenge. The thermal noise (one of the most fundamental limits for any instrument) can be lowered significantly with increasing observing time. However, other noise sources like faint unmodelled sources \citep{ewall2017}, ionospheric variation, instrument response, etc., cannot be reduced by increasing integration time. Thus accurate calibration can increase SNR, but up to a certain limit as allowed by observing conditions. This work quantifies the actual level of calibration accuracy required for upcoming radio interferometers to obtain "good" SNR levels.

The bright radio sources that contaminate the data sets are expected to be removed before extracting the cosmological 21-cm signal.  While the predominant source of the bright foreground is the diffuse galactic emission, the point source contamination also poses a problem. The problem arises mainly because modeling and removal of each source is challenging. Any calibration imperfections give rise to artifacts in the residual that may affect the signal extraction adversely \citep{Datta2009, Datta2010}. Moreover, these observations have limited resolution due to the signal being coherent at short baselines only. Thus, high-resolution observations are required for accurate modeling and removal of discrete foregrounds. Hence, there are very stringent accuracy requirements for the sky model. Further calibration cannot minimize any error inherent to such a model, and the same will result in contaminated residuals. Thus position errors pose a challenge for the detection of the cosmological signal.
Moreover, due to phase errors from directional effects, source positions may also shift. This will result in incorrect point source removal resulting in artifacts affecting signal extraction. Most sources have fluxes $\sim$ mJy or lower at the target frequencies for real observations. Since the cosmological signal is feeble, compact sources with even such "low" fluxes remaining in the residuals pose a problem. It has been shown that while redundant calibration is potentially better for arrays with a high degree of redundancy like HERA and the MWA Hexagonal configuration, offsets in source positions systematically affect the phase solutions \citep{Joseph2018}. A global sky model (having only compact sources) as observed in actual observation is considered for this work. The effect of errors in the source position through a systematic error in the right ascension of the sources is investigated here.

\section{Formalism for Analyses}
\label{formalism}

\begin{figure}
\centering
    \includegraphics[width=6.1cm, height=4.8cm]{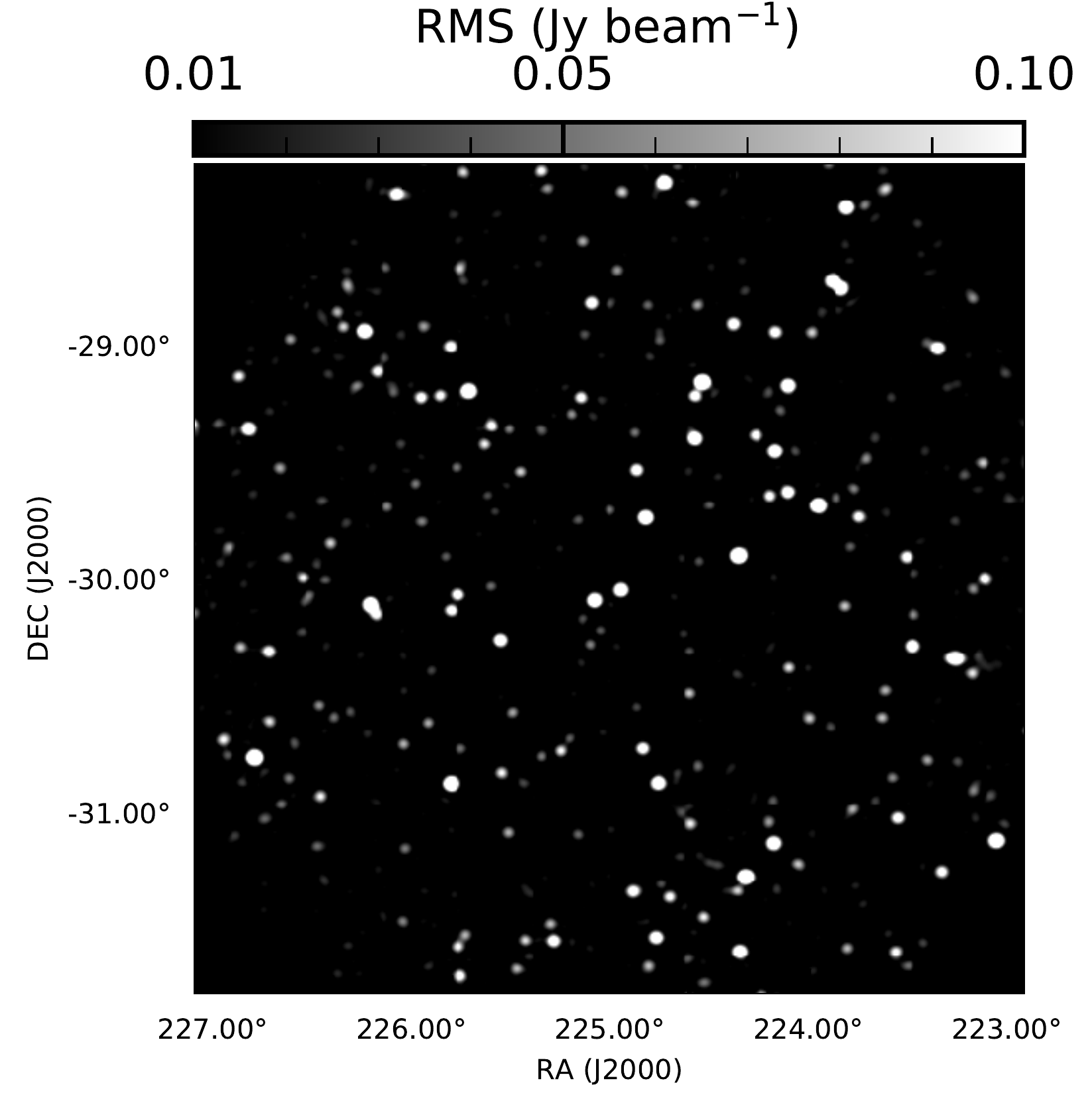}
    \includegraphics[width=6cm, height=4.8cm]{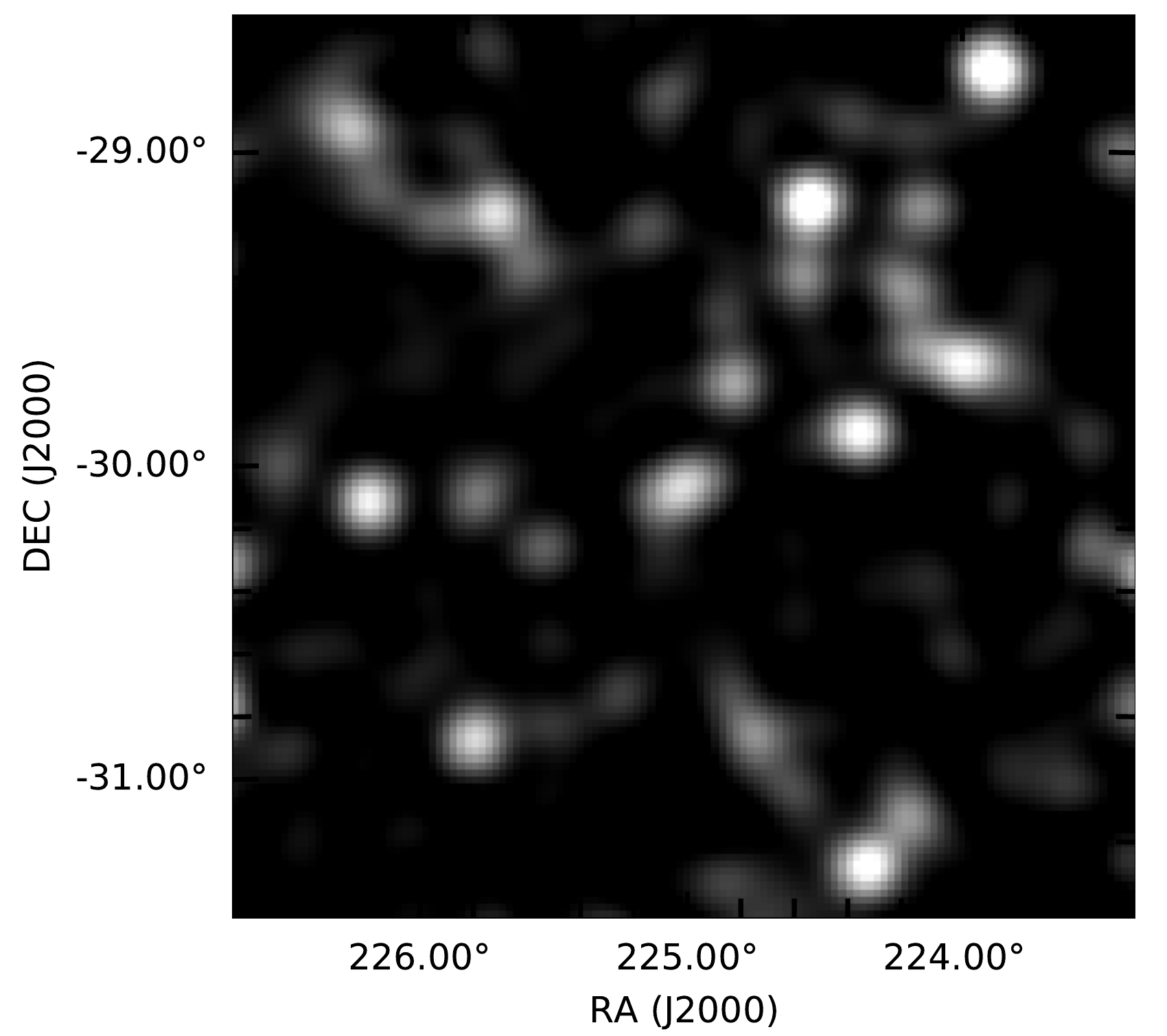}
    \includegraphics[width=6cm, height=4.8cm]{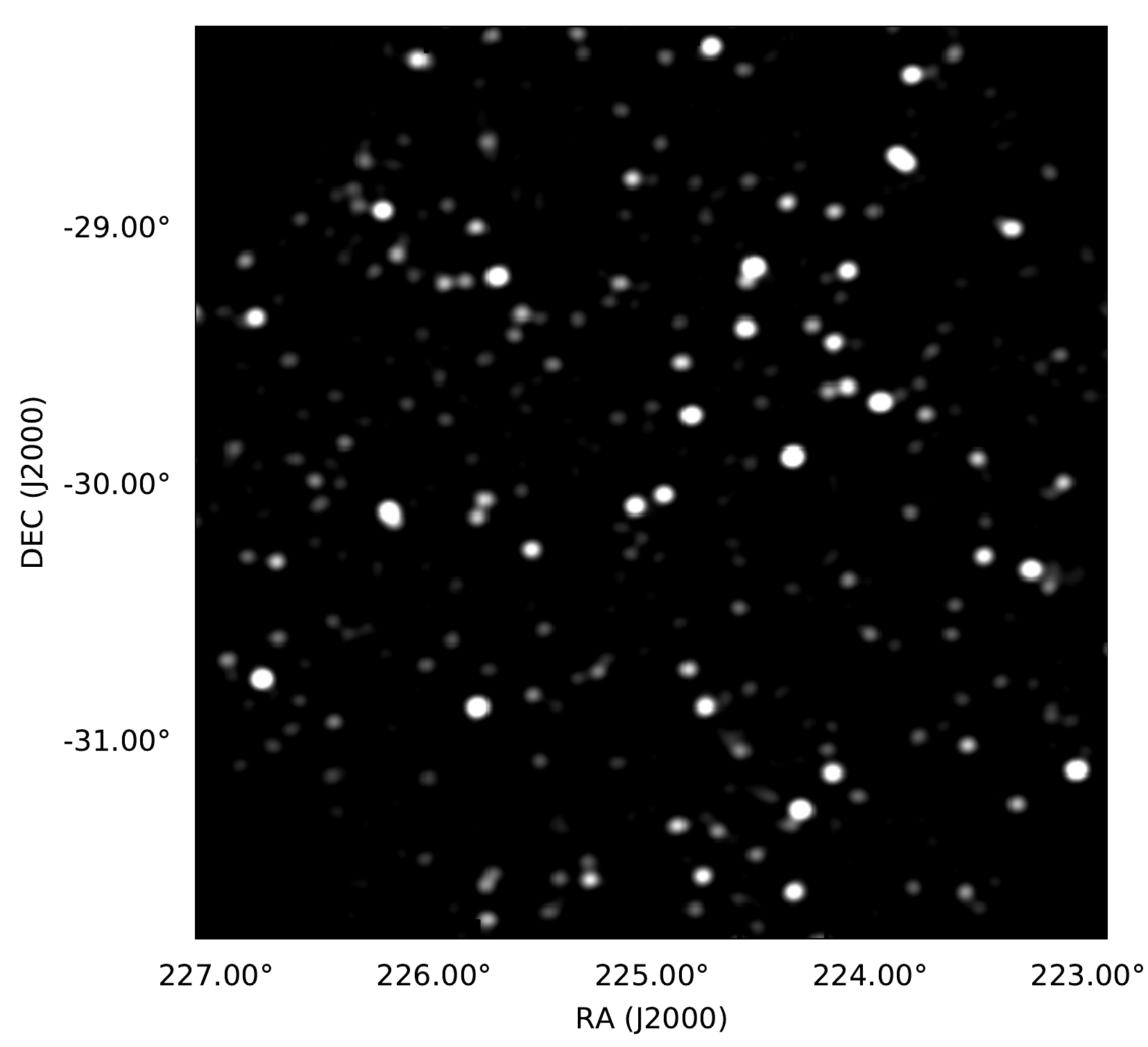}
    \includegraphics[width=6.25cm, height=4.8cm]{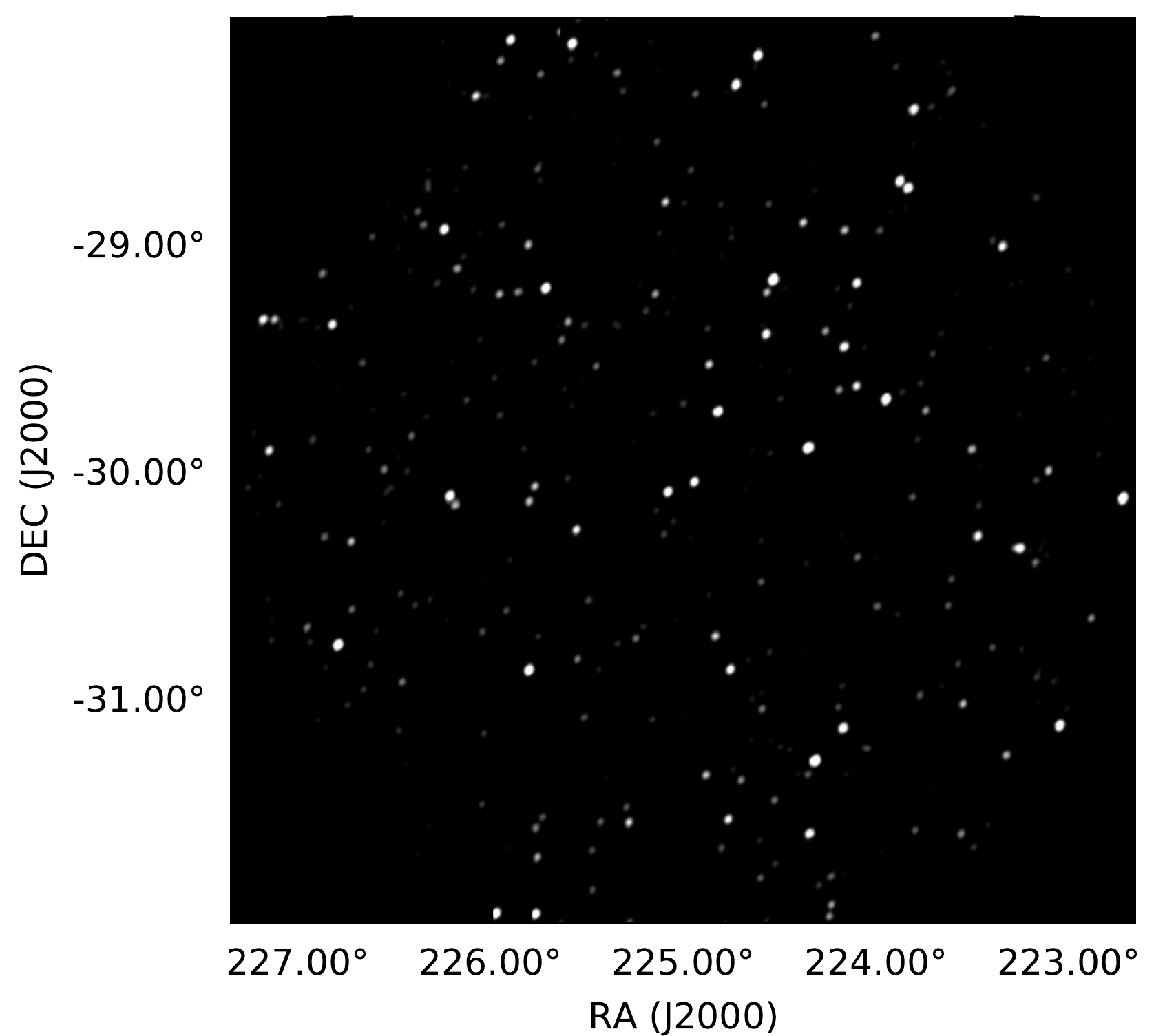}
    \caption{Simulated sky with T-RECS as foreground model for (top to bottom) SKA1-Low, HERA, MWA-I \& MWA-256.}
    \label{sim_sky}
\end{figure}

\begin{figure}
    \includegraphics[width=\columnwidth, height=6cm]{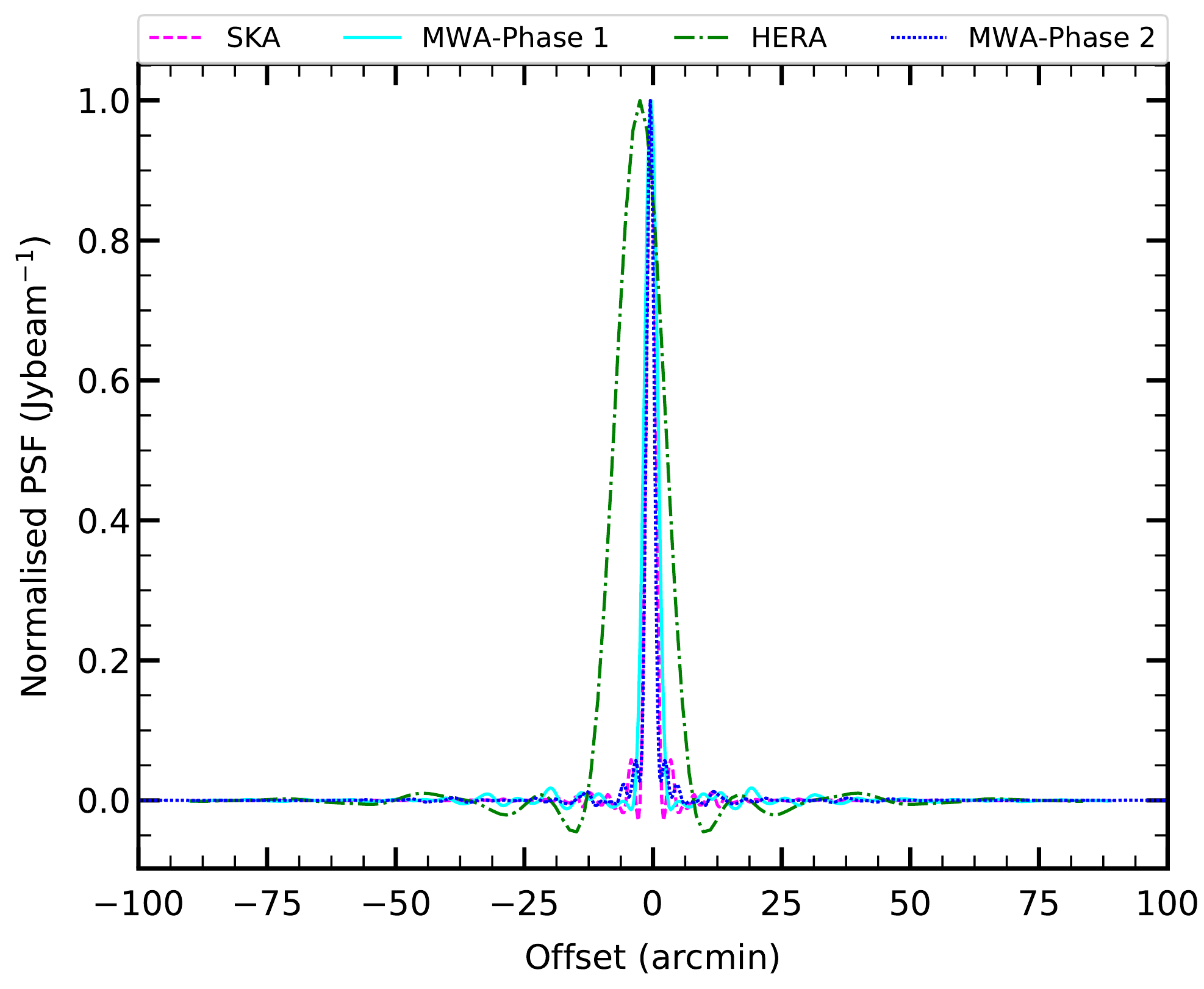}
    \caption{Normalized Point Spread Function using robust parameter 0 at the central frequency of observation for SKA1-Low(magenta), HERA(green), MWA-1(cyan) and MWA-2(blue).}
    \label{psf}
\end{figure}

The visibilities obtained from the synthetic observations described in Section \ref{Simulation} are used for further analysis. Figure \ref{sim_sky} shows the simulated image of the sky with T-RECS as foreground model for SKA1-Low, HERA, MWA-I \& II respectively (top to bottom respectively). Figure \ref{psf} shows the point spread function (PSF) at the central frequency of observation for each of the telescopes used. It can be seen from the figure that the best PSF response comes from SKA1-Low (magenta curve), both in terms of width and amplitude of side lobes. PSF for HERA (shown by green curve) has the widest response, which is, because it has the shortest maximum baseline. However, due to the layout and configuration of the array, it has a small side-lobe amplitude. The MWA configurations (Phase-1 in cyan and Phase-2 in blue) show high amplitude oscillatory features in the sidelobes.

For determination of the effect of calibration errors, the ideal visibilities are corrupted by complex gains (as described in Equation \ref{me}) . The errors in gain phases and amplitudes (see Equation \ref{gaineq}) were drawn from a 0 mean Gaussian distribution with a standard deviation equal to the percentage of error introduced. For this work, 0.001\%, 0.01\%, 0.1\%, 1.0\% \& 10.0\% gain errors are introduced in both the gain phases and amplitudes. The model visibilities are subtracted from the corrupted ones to produce residual visibilities, i.e.,
\begin{equation}
    V{_{ij}^{residual}} = V{_{ij}^{observed}} - V{_{ij}^{model}} 
    \label{res_eq}
\end{equation}
Here, $V{_{ij}^{observed}}$ is the corrupted visibility (the label 'observed' implies actual observations the instrument records where the visibilities come with various imperfections). 

The effect of position errors (i.e., in the global sky model) is determined by simulating observations with erroneous sky models. These inaccurate sky models are generated, taking the PSF size of HERA at the observing frequency (enlisted in Table \ref{paramters}) as a metric to quantify the offsets. A zero mean Gaussian distribution with standard different deviations- 0.01\%, 0.1\%, 1.0\%, 10.0\% \& 100.0\% of 8.1\arcmin (486\arcsec) \footnote{This means that the maximum displacement allowed are $\sim$ 0.05\arcsec, 0.5\arcsec, 5.0\arcsec, 50\arcsec and 500\arcsec respectively for the percentages of error mentioned.} are used to generate different sets of erroneous sky models.
The position errors thus generated were added to the RA of the sources to create a new catalog with inherent errors in the source positions. The residuals are obtained similarly, as discussed for Equation \ref{res_eq}. 
Residuals obtained by subtracting the corrupted sky from the true one have been used to determine the effect in both the image plane and the PS. The formalism used to quantify the image plane effects and PS is described briefly in the following subsections. It should be mentioned that the position error in only the RA has been introduced for simplicity. Such errors can be present in either one of RA or DEC or both in a point source model. Such errors will obscure the EoR signal beyond a certain offset.

\subsection{Image Plane}
While most EoR detection experiments do not target imaging due to low signal-to-noise ratio (SNR), the SKA1-Low is projected to produce tomographic images with SNR$\gtrapprox$1 \citep{Mellema2014}. There are various techniques, including machine learning approaches and the traditional \textit{CLEAN} \citep{CLEAN1974} algorithm that is being explored for maximum information extraction in the image domain (for instance, see \citet{Liu_shaw_2020} for a comprehensive summary). However, the efficacy and robustness of these methods are not determined for a wide variety of actual observations.  Thus it is worthwhile to investigate how the errors that have been introduced here affect the image plane. The analysis is done by constructing a simple dirty map in CASA and using the RMS in the image plane as the quantifying metric (note that a comparison of various techniques for image recovery has not been made since it is beyond the scope of the current work). The pure \hi\ signal used in the input has a peak flux of $\sim$ 0.8 $\mathrm{\mu Jy beam^{-1}}$ at 142 MHz, hence to get a significant detection, the residual RMS should lie below this level. Thus, performance in terms of the RMS has been compared. Additionally, image plane performance is also quantifiable using its dynamic range (DR) - the ratio of the peak flux in the image and the residual RMS. In the case of the presence of imperfect calibration, the theoretical DR is defined as \citet{Taylor1999} : 
\begin{equation}
   DR = \sqrt{\frac{N(N-1)}{2(a^2+\phi^2)}}
    \label{dr_eq}
\end{equation}
with a and $\phi$ as amplitude and phase, respectively, and N is the number of interferometer elements. The lowest RMS obtained from the residual image near the phase center is used to quantify the effect of the errors mentioned earlier. Since two different foreground models are used for this work, they are represented via the dashed (EN1) and dot-dashed (T-RECS) curves.

\begin{figure}
    \includegraphics[width=\columnwidth,height=7cm]{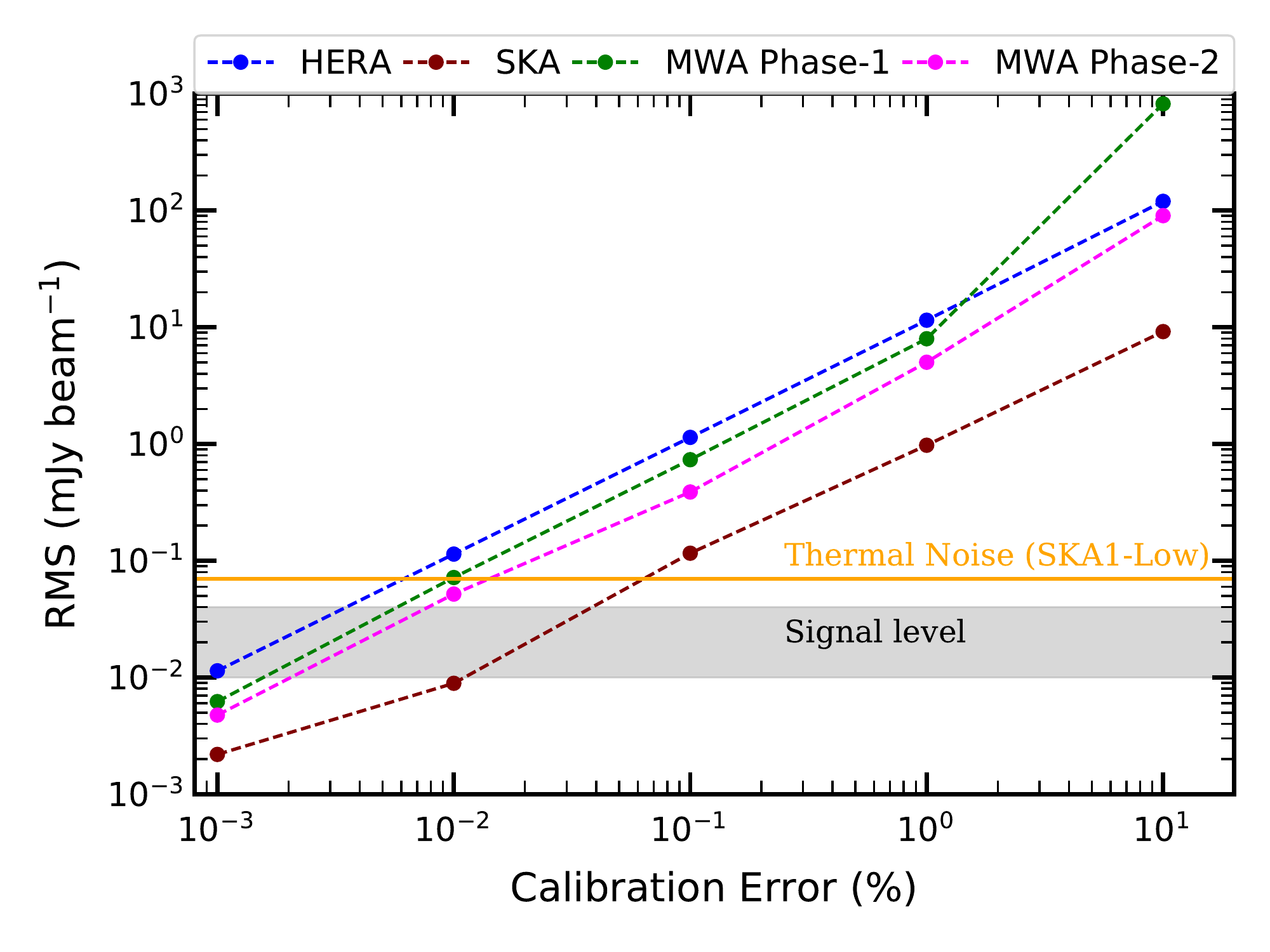}
    \caption{ Variation of RMS in the residual image as a percentage of calibration error for T-RECS catalog as foreground model. The grey region shows the amplitude of range of observed signal. The solid orange line is the thermal noise limit for a 4-hour observation with SKA1-Low.}
    \label{gain_rms}
\end{figure}

\begin{figure}
    \includegraphics[width=\columnwidth,height=3in]{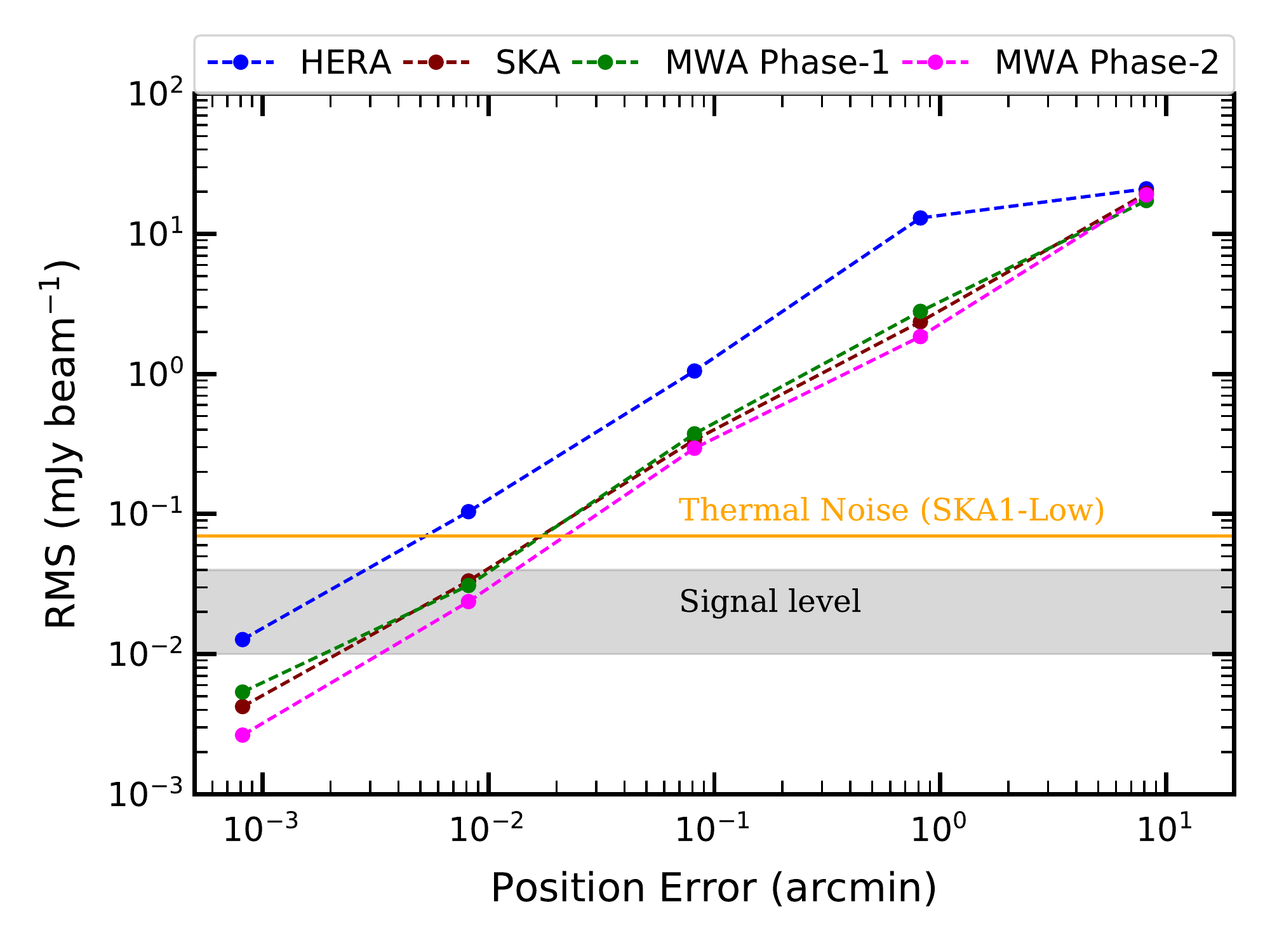}
    \caption{Variation of RMS in residual image with position error using T-RECS as foreground model for the four telescope configurations considered. The RMS is plotted as a function of the maximum displacement of the sources (in arcmin) from the true position. The solid orange line is the thermal noise limit for a 4-hour observation with SKA1-Low.}
    \label{pos_rms}
\end{figure}

Figure \ref{gain_rms} summarises the results obtained from the gain residuals. For the dynamic range, the general trend is consistent with theoretical expectations. The desirable calibration accuracy obtained from these simulations translates to a dynamic range requirement of $\sim$10$^{5}$ or higher, which in turn implies the necessity of a highly accurate gain solution for observations with the SKA1-Low. Figure \ref{pos_rms} shows the variation of RMS in the residuals with position error as a function of angular size. In both Figures \ref{gain_rms} and \ref{pos_rms}, the gray band represents the maximum and minimum signal amplitudes in the image plane.

In the best case scenario, the sensitivity of the instrument will be determined by the thermal noise in the system. The thermal noise limit is given by \citep{fan2006}:
\begin{equation}
   \sigma_{T} = \Big(\frac{1.9}{\sqrt{\Delta\nu_{kHz}t_{hr}}}\Big)\Big(\frac{T_{sys}}{A_{eff}N_{ant}}\Big)
\end{equation}
where $\sigma_{T}$ is the thermal noise in the system, $\Delta\nu_{kHz}$ is the channel width in kHz, $t_{hr}$ is the integration time in hours, $T_{sys}$ is the system temperature, $A_{eff}$ is the antenna collecting area and $N_{ant}$ is the number of antennas. At the frequencies of interest, $T_{sys}$ is equivalent to the sky temperature. The sky temperature, $T_{sky}$ is calculated using $T_{sky} \sim 180\Big(\frac{\nu}{180 MHz}\Big)^{-2.6}$K \citep{Furlanetto2006}. The values of thermal noise is tabulated in Table \ref{paramters}. As can be seen, the least thermal noise level for a 4 hour observation is obtained for SKA1-Low, which is shown via the solid orange line in Figures \ref{gain_rms} \& \ref{pos_rms}. In the case of SKA1-Low, \citet{ska_science} states that the spectral line sensitivity at the frequency band of interest is 1258 $\mu$Jy beam$^{-1}$ for a fractional bandwidth of 10$^{-4}$, and integration time $\Delta\tau$ of 1 hour; for the specifications used in this work, the sensitivity comes down to $\sim$220 $\mu$Jy beam$^{-1}$.

\subsection{Power Spectrum}
Statistical detection of the redshifted cosmological 21-cm signal is the primary target of most low-frequency interferometric arrays. For EoR experiments, there is a formalism to estimate the PS from both visibility (for instance, see \citealt{mwa2}) as well as from the image plane (for example \citealt{lofar2}); this work uses visibility for its determination. The interferometric visibility is the correlation between the signals received between two antenna pairs, which is given by \citep{Taylor1999} :
\begin{equation}
    V ({\mathbf{U}},\nu) = \iint A(\hat{\mathbf{s}},\nu) B(\nu) I(\hat{\mathbf{s}},\nu) e^{-i2\pi \nu \mathbf{U}.\hat{\mathbf{s}}} d\Omega,
    \label{vis_eq}
\end{equation}
where, ${\mathbf{U}}$ is the baseline vector, $I(\hat{\mathbf{s}},\nu)$ and $ A(\hat{\mathbf{s}},\nu)$ are the specific intensity and antenna beam pattern as a function of frequency ($\nu$) and  $B(\nu)$ is the instrumental bandpass response; the unit vector as,  $\hat{\mathbf{s}} \equiv (l,m,n)$, where $l,m,n$ are the direction cosines towards east, north and zenith respectively with $n = \sqrt{1-l^{2}-m^{2}}$ and $ d\Omega = \frac{dl dm}{\sqrt{1-l^{2}-m^{2}}}$. For this work, $ A(\hat{\mathbf{s}},\nu)$ is taken to be 1, i.e. the effect of primary beam is not considered. 

Inverse Fourier transform of $V ({\mathbf{U}},\nu)$ along the the frequency axis gives the visibility in the delay domain (henceforth represented as $\tau$), $V ({\mathbf{U}},\tau)$. Using this formalism, the cylindrical PS (as discussed in \citealt{Morales2004}) is given by:

\begin{equation}
    P(\mathbf{k}_{\perp}, k_{\parallel}) = \Big(\frac{\lambda^{2}}{2k_{B}}\Big)^{2}    \Big(\frac{X^{2}Y}{\Omega B}\Big)  |V (\mathbf{U},\tau)|^{2}, 
    \label{2dps_eq}
\end{equation}
with unit $K^2(Mpc/h)^3$. In the above equation, $\lambda$ is the wavelength corresponding to central frequency (2.1 m for these simulations), $k_{B}$ is the Boltzmann constant, $\Omega$ is the primary beam response, B is the bandwidth (8 MHz for this work), X and Y are the conversion
factors from angle and frequency to transverse co-moving distance (D(z))  and the co-moving depth along the line of sight, respectively \citep{Morales2004}. The quantity $\mathbf{k}_{\perp}$ represents the Fourier modes perpendicular \& $k_{\parallel}$ represent the k modes along the line of sight, given by 
 \begin{equation*}
 \mathbf{k}_{\perp} = \frac{2\pi |\mathbf{U}|}{D(z)} \hspace{20pt} \& \hspace{20pt}
  k_{\parallel} = \frac{2\pi \tau \nu_{21} H_{0} E(z)}{c(1+z)^{2}}
 \end{equation*}
where $\nu_{21}$ is the rest-frame frequency of the 21 cm spin flip transition of \hi\, $z$ is the redshift corresponding to the observing frequency, $H_{0}$ is the Hubble parameter and $E(z) \equiv [\Omega_{\mathrm{M}}(1+\textit{z})^{3} +  \Omega_{\Lambda}]^{1/2}$.  $\Omega_{\mathrm{M}}$ and $\Omega_{\Lambda}$ are matter and dark energy densities, respectively \citep{Hogg1999}. 

From the 2D power spectrum, the 1D PS (obtained by spherical averaging of $P(\mathbf{k}_{\perp}, k_{\parallel})$)is calculated as :
\begin{equation}
   \Delta^{2} (k) = \frac{K^{3}}{2\pi^{2}} \langle P(\mathbf{k})\rangle _{k}
   \label{3d_ps}
\end{equation}
where, $k = \sqrt{k_{\perp}^{2} + k_{\parallel}^{2}}$. 
The uncertainty in PS has been calculated using a modified version of 21cmSense \citep{baobab} for each array. The 3$\rm \sigma$ error bars have been put in the spherically averaged PS to show the uncertainty in power obtained. 

In the case of SKA1-Low, it is still not decided whether signal PS estimation will be done using foreground removal, avoidance, or some other technique. This work uses the generic method of averaging over all k-modes to determine the 1D power. The same approach is also used for both MWA configurations and HERA to maintain consistency (although in \citealt{mwa2} and \citealt{2021arXiv210802263T}, the upper limits have been placed using avoidance methodology).

%%%%%%%%%%%%%%%%%%%%%%%%%%%%%%%%%%%%%%%%%%%%%%%%%%%%%%%%%%%%%%%%%%%%%%%%%%%5
%%%%%%%%%%%%%%%%%%%%%%%%%%%%%%%%%%%%%%%%%%%%%%%%%%%%%%%%%%%%%%%%%%%%%%%%%%%%

%%%%%%%%%%%%%%%%%%%%%%%%%%%%%%%%%%%%%%%%%%%%%%%%%%%%%%%%%%%%%%%%%%%%%%
%%%%%%%%%%%%%%%%%%%%%%%%%%%%%%%%%%%%%%%%%%%%%%%%%%%%%%%%%%%%%%%%%%%%%%%%%%%

The results obtained for each array configuration for imaging and PS estimation are described in the following section. It should be mentioned that both the foreground models give consistent results for all the telescope configurations considered. The power spectrum performance plots shown in the following subsections use T-RECS as the foreground model. A representative case using EN1 as foreground model has been shown in the appendix for all the configurations to emphasize that the pipeline performs consistently irrespective of the sky model used.

\section{Results}
\label{results}
This section describes the results obtained from the simulations and discusses their implications for actual observations. The results for each of the configuration considered is described in separate subsections below. 

\subsection{SKA1-Low}

The synthetic observation pipeline is primarily being developed to simulate the effects of non-ideal conditions that are present in actual observations. The primary motivation is to determine how these errors will affect the performance of SKA1-Low for the sensitive low-frequency observation done for CD/EoR science. The following subsections describe the effect of calibration and position errors for SKA1-Low.

\subsubsection{Calibration Error}
The maroon curve shows the effect of calibration error on SKA1-Low in the image plane in Figure \ref{gain_rms}. It is seen that for calibration inaccuracy $\sim$10$^{-2}$\%, the residual RMS is well below the signal level. 0.01\% represents the borderline case where the RMS and thermal noise are nearly similar. However, the residual RMS level exceeds the signal level by almost an order of magnitude, indicating that the limiting factor for these observations would be systematics (instead of thermal noise) for sufficiently large integration times using the SKA1-Low. For even higher errors, the image RMS is high enough to confuse or even obscure the faint cosmological signal detection. 

%%%%%%%%%%%%%%%%%%%%%%%%%%%%%%%%%%%%%%%%%%%%%%%%%%
 \begin{figure*}
    \includegraphics[width=\columnwidth, height=8cm]{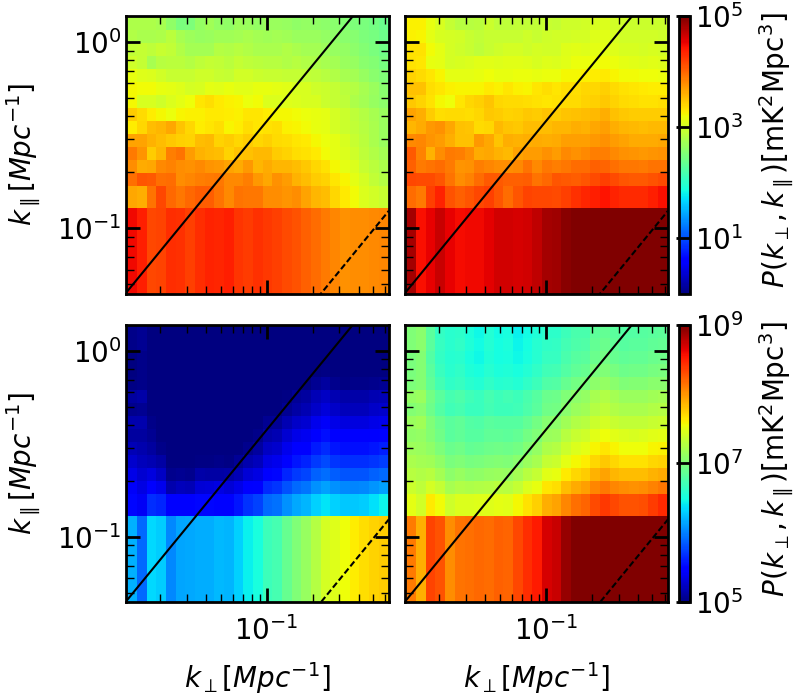}
   \includegraphics[width=\columnwidth,height=8cm]{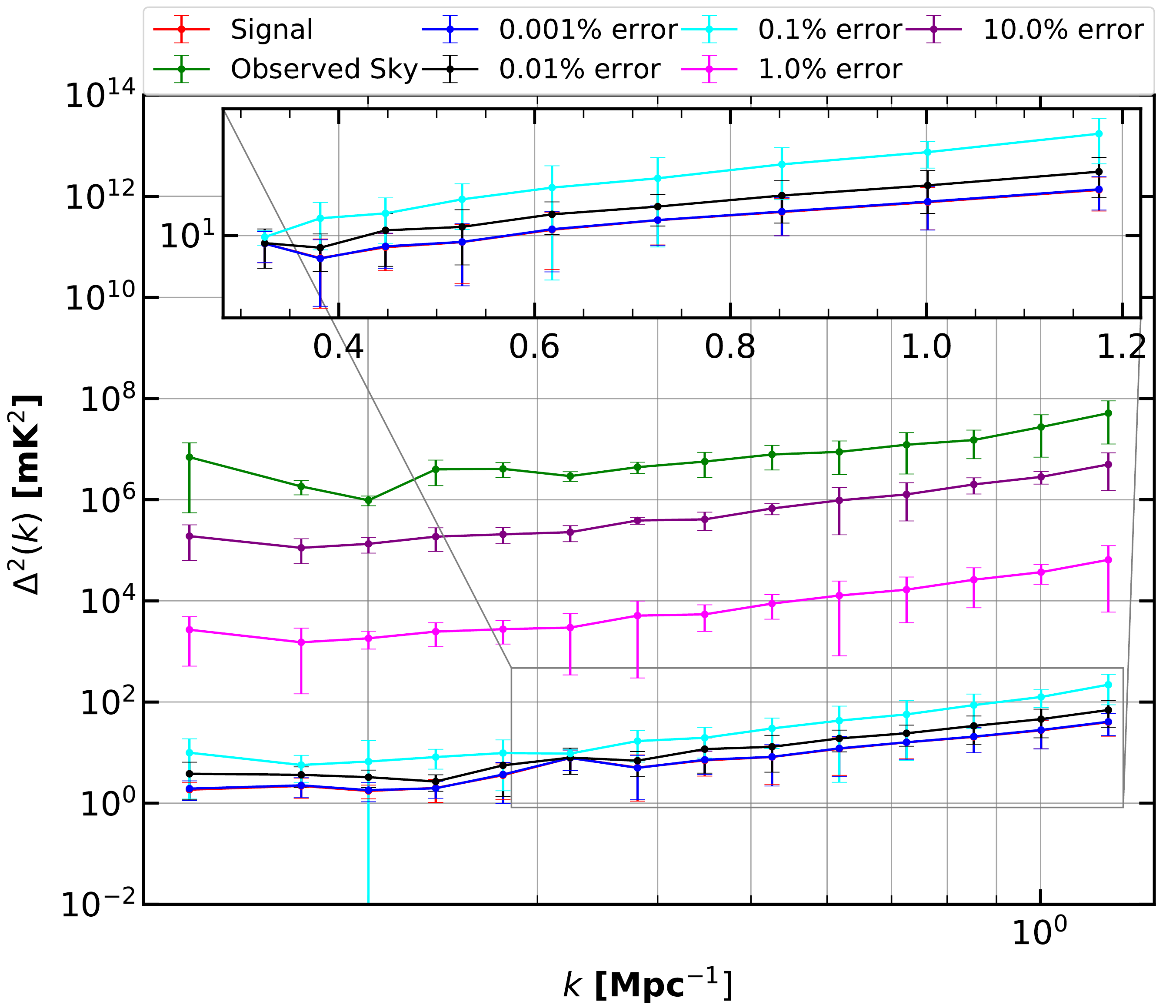}  
\caption{Residual power spectrum for gain errors with SKA-1 Low. \textbf{(left)} 2D PS of the  residual visibilities for calibration errors for 0.001\% (top-left), 0.01\% (top-right), 0.1\% (bottom-left) \& 1.0 \% (bottom-right). The solid black line represents the horizon line, while the dashed black line is the FoV limit. \textbf{(right)} Spherically averaged PS with calibration residuals compared with the signal power and foreground power. The error bars are 3$\rm \sigma$ uncertainties for the k-bins including sample variance and thermal noise.}
     \label{ska_g}
 \end{figure*}
%%%%%%%%%%%%%%%%%%%%%%%%%%%%%%%%%%%%%%%%%%%  

In Figure \ref{ska_g}, the left panel shows the 2D power spectrum of the residual visibilities for SKA1-Low. It is seen that up to 0.01\% calibration error, the wedge stays $k_{\perp}$ modes below $\sim$ 0.1 and below $k_{\parallel} \sim$ 0.2, and the EoR window region have lower contamination. For cases exceeding 0.1\% error, the EoR window has a very high power amplitude, implying the residual foregrounds dominate above inaccuracies $\sim$ 0.1\% in gain amplitude. 

The right panel of Figure \ref{ska_g} shows the 1-dimensional spherically averaged power spectrum for gain error residuals. The zoomed-in part shows the region near the signal power at k $\sim$0.2 and above. The green curve shows the power for the observed sky (i.e., the sky with both signal and foregrounds). It is seen that at 0.001\% error (blue curve), the residual power overlaps with the signal power (red curve).  At 0.01\%, the power spectrum starts deviation from the signal level at k=0.5 Mpc$^{-1}$ and above, but are still consistent within error-bars. But for 1\% and above errors in calibration, the foreground contribution in the residuals overrides the signal power completely. Thus, if all other sources of error are well understood and mitigated, the residual calibration error can render the signal obscured if the algorithm accuracy is $\sim$1\% or worse.

\subsubsection{Position Error}

The maroon curves show the effect of position error in the image plane in Figure \ref{pos_rms}. The top panel shows the RMS as a function of maximum displacement from the actual position, and the bottom curve represents the same as a function of the relative percentage of error. For errors up to $\sim$ 0.048\arcsec (0.1\% of the PSF size), the residual RMS stays well below the signal level, thereby making its detection feasible. At 1\% error (or 0.48\arcsec displacement), the RMS residual RMS starts getting dominated by the residual foregrounds, which again can potentially confuse detection of the redshifted signal. It is also seen that up to a maximum displacement of $\sim$ 0.048\arcsec, the error remains below the thermal noise level for 4-hour observation (orange line), again emphasizing for a sufficiently large integration time, observations would be systematics limited rather than noise limited.

 \begin{figure*}
    \includegraphics[width=\columnwidth, height=8cm]{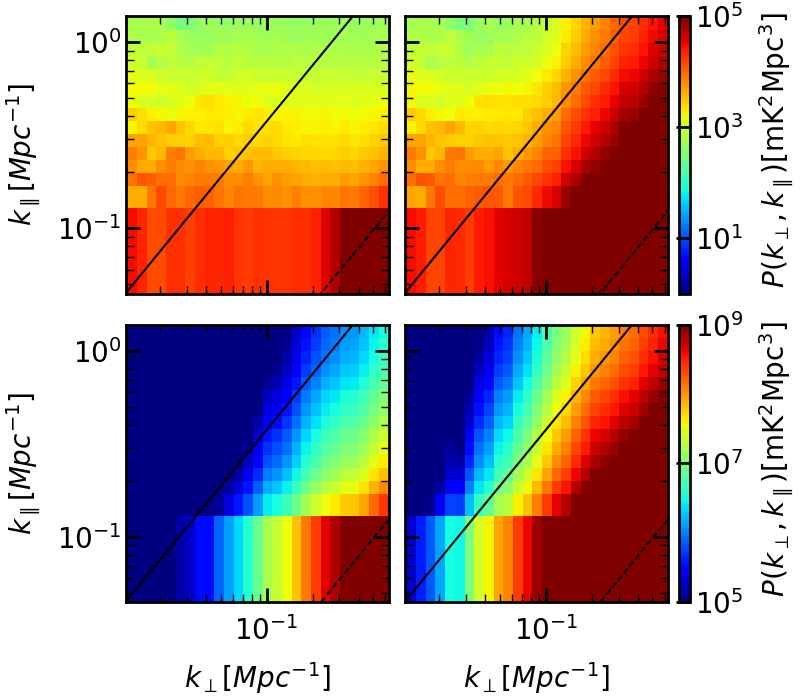}
        \includegraphics[width=\columnwidth, height=8cm]{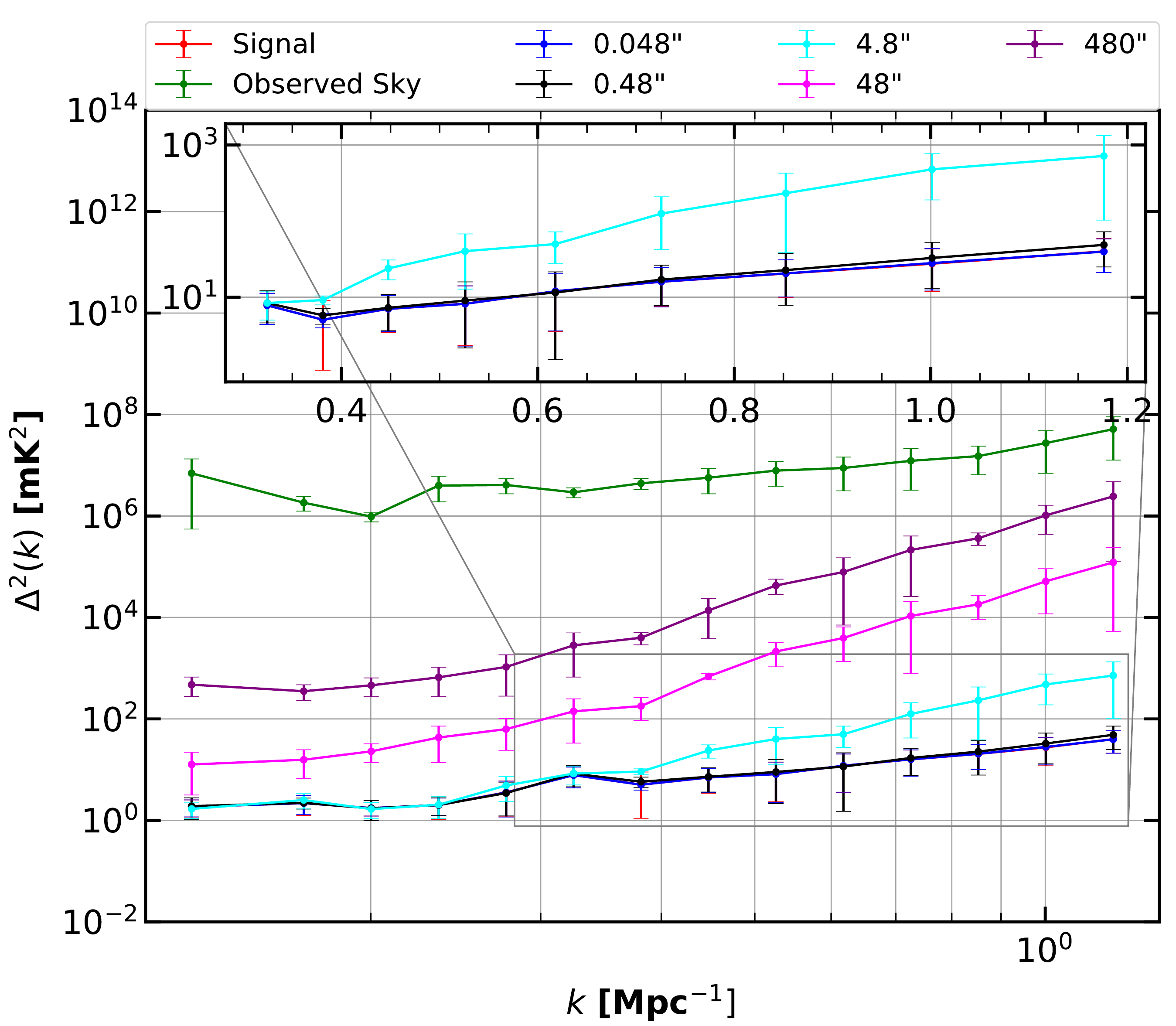}
\caption{Residual power spectrum for position errors with SKA1-Low. \textbf{(left)} 2D cylindrical averaged power spectra of the  residual visibilities for position errors with 0.1\% (top-left), 1\% (top-right), 10\% (bottom-left) \& 100 \% (bottom-right). The solid black line represents the horizon line, while the dashed black line is the FoV limit. \textbf{(right)} Spherically averaged PS with position residuals compared with the signal power and foreground power.The error bars are 3$\rm \sigma$ uncertainties for the k-bins including sample variance and thermal noise.}
\label{ska_p}
 \end{figure*}
%%%%%%%%%%%%%%%%%%%%%%%%%%%%%%%%%%%%%%%%%%%%%%%%%%%%%%%%%%%%%%%%%

In Figure \ref{ska_p}, the residual power for position errors with SKA1-Low is shown. The left panel shows the 2D PS, while the right panel 1D spherically averaged PS. It is seen for the residual 2D power spectrum that with offsets $\geq$ 10\% of PSF size (i.e., 40\arcsec and above), the residual foregrounds cause leakage into the k-modes outside the wedge, giving a significantly high amplitude of power. Subsequently, in the case of the 1D PS, the residual power overrides the signal at all k modes by orders of magnitude. This is also reflected in the 1D power spectrum.  For a position error of $\sim$0.1\% (i.e., 4\arcsec), the signal and the residual foreground powers are at par with each other. Thus, it can be seen from both the figures that the position accuracy required is $\sim$0.01\arcmin to render the signal detectable. Hence in the case of PS estimation, it can be seen that for achievable positional accuracy at the observing frequency, the statistical detection of the cosmological signal is possible.

\subsection{HERA}

HERA is one of the key EoR instruments that has already begun its observation and has recently provided its first upper limit using the data from the initial observing phase using 52 antennas \citep{2021arXiv210802263T}. For this work, the entire 350 antenna configuration (see Figure \ref{arrays}, top right panel) has been used. The effect of errors seen in both image and PS domain is described below.

\subsubsection{Calibration Errors}
The blue curves in Figure \ref{gain_rms} show the effect of calibration residuals in the image plane for HERA. For both the sky models, it is seen that the residual RMS for gain solutions that have as small as 0.01\% and above errors exceeds the signal power level. Thus, under identical observing conditions and in the presence of similar errors, SKA1-Low outperforms HERA in terms of calibration error tolerance by an order of magnitude in the image plane. 

\begin{figure*}
    \includegraphics[width=\columnwidth, height=8cm]{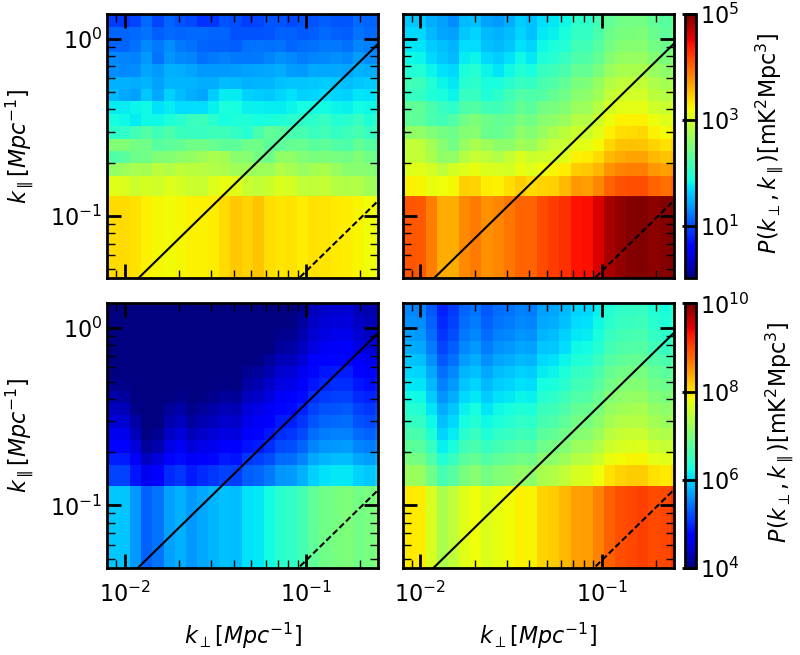}
      \includegraphics[width=\columnwidth, height=8cm]{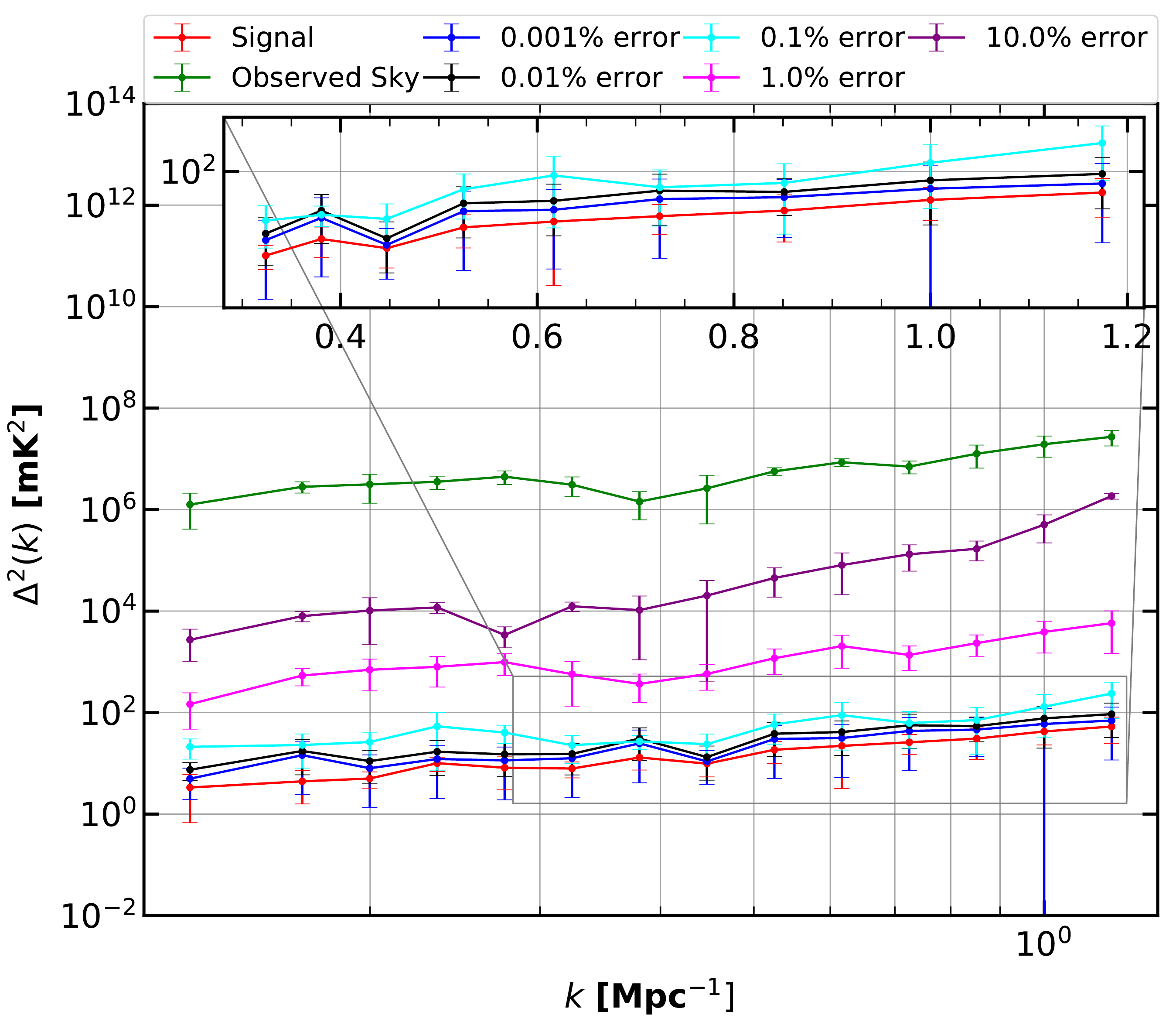}
     \caption{Residual power spectrum for gain errors with HERA. \textbf{(left)} 2D PS of the  residual visibilities for calibration errors for 0.001\% (top-left), 0.01\% (top-right), 0.1\% (bottom-left) \& 1.0 \% (bottom-right). The solid black line represents the horizon line, while the dashed black line is the FoV limit. \textbf{(right)} Spherically averaged PS with calibration residuals compared with the signal power and foreground power.The error bars are 3$\rm \sigma$ uncertainties for the k-bins including sample variance and thermal noise.}
     
  \label{hera_g}
 \end{figure*}

The PS for HERA for gain error residuals is shown in Figure \ref{hera_g}. The left panel is the 2D PS, while the right panel is the 1D PS. In the $k_{\parallel}$- $k_{\perp}$ plane, the wedge can be observed in the modes within $k_{\perp}\sim$ 0.01 and 0.3 and $k_{\parallel}\geq $ 0.3 are free from very high amplitude of foregrounds.

The residual 2D PS shows that for calibration errors above $\sim$0.1\%, the foregrounds power becomes high and starts to dominate at all k modes. This is also evident from the 1D power spectrum of the calibration residuals (Figure \ref{hera_g} right panel), where it is observed that upto 0.1\%, the residual power very slightly deviates from the signal power. However, for inaccuracies $\geq$0.1\%, the residual power exceeds the signal by a significant factor, overruling the signal and obscuring it.

\subsubsection{Position Errors}

In Figure \ref{pos_rms}, the blue curves represent the residuals of calibration error for synthetic observations with HERA. In this case, an offset of 0.4\arcsec makes the residual RMS low enough to detect the signal. Hence, for HERA, offsets less than 0.4\arcsec are required for the image RMS to lie below the signal level.

\begin{figure*}
 \includegraphics[width=\columnwidth, height=8cm]{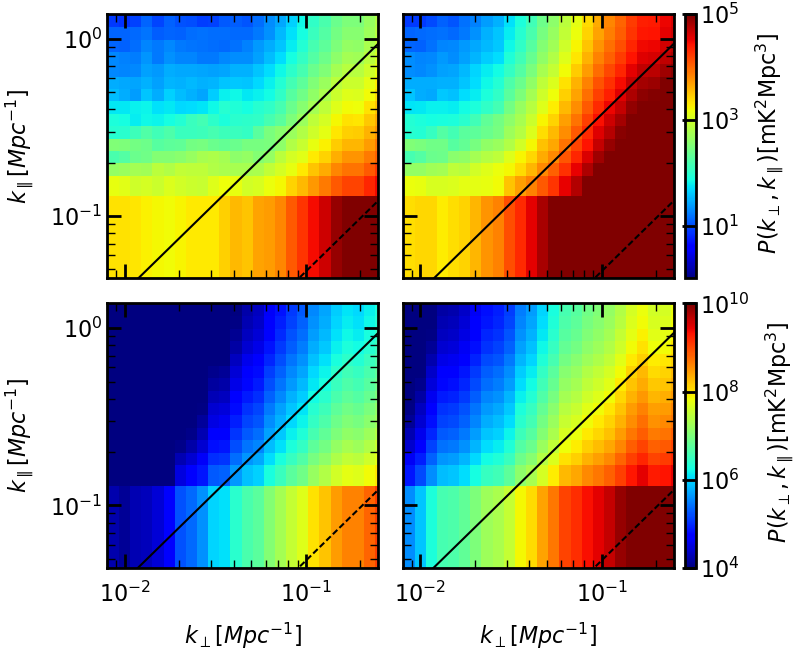}
    \includegraphics[width=\columnwidth, height=8cm]{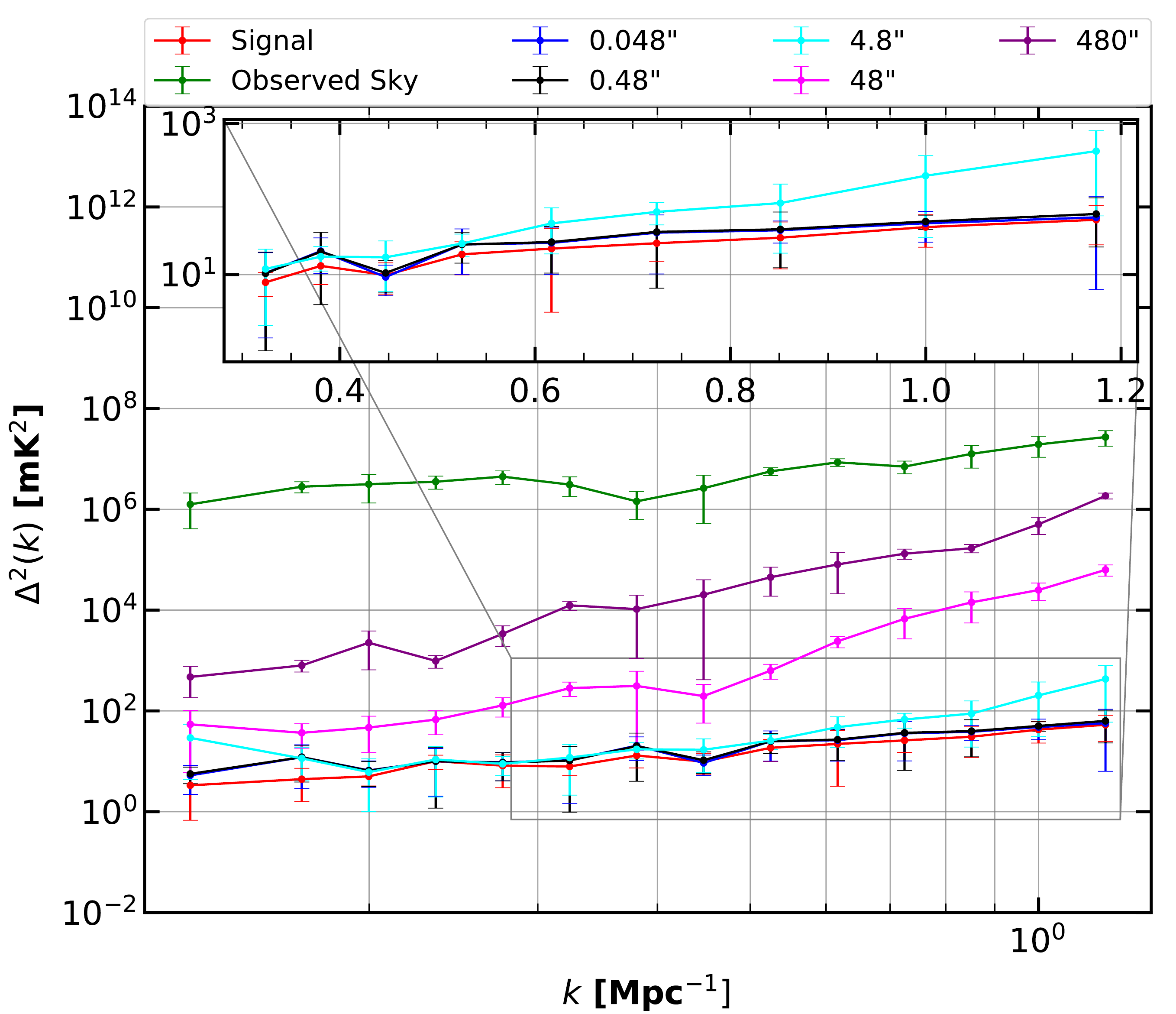}
\caption{Residual power spectrum for position errors with HERA. \textbf{(left)} 2D cylindrical averaged power spectra of the  residual visibilities for position errors with 0.1\% (top-left), 1\% (top-right), 10\% (bottom-left) \& 100 \% (bottom-right). The solid black line represents the horizon line, while the dashed black line is the FoV limit. \textbf{(right)} Spherically averaged PS with position residuals compared with the signal power and foreground power.The error bars are 3$\rm \sigma$ uncertainties for the k-bins including sample variance and thermal noise.}
 \label{hera_p}
  \end{figure*}

The residual PS for position error residuals for HERA is shown in Figure \ref{hera_p}. It is observed from the left panel that for deviations of $>$1\% of the "actual" position, the power from the residual foregrounds contaminates most of the k-modes, causing the signal to be obscured. A similar trend is seen in the right panel, where the signal power spectrum (red curve) is being followed (with slight deviations of a factor of $\sim$1.3) for position errors up to $\sim$0.01\arcsec (0.1\% of PSF size). At 1\% of the PSF size ($\sim$0.1\arcsec), the residual power deviates from the signal power at k modes above 0.5 Mpc $^{-1}$. Above 1\% error, the deviations are orders of magnitude higher, which results in the foreground residuals obscuring the signal.

\subsection{MWA}

The MWA is one of the precursor facilities to the SKA1-Low. In the initial phase of operation, it consisted of 128 square tiles extended up to $\sim$3 km. The second phase of the MWA, after its upgrade, consists of 256 tiles, which spreads out to $\sim$5 km. Both the old and the upgraded configurations with all the tiles have been used for this work. The results for the introduction of the errors as mentioned earlier are discussed in the following subsections.

\subsubsection{Calibration Error}

Green and magenta curves in Figure \ref{gain_rms} show the performance of MWA-I and MWA-256 respectively in the image domain in the presence of residual calibration errors. For both the cases, RMS corresponding to residuals of 0.01\% is below the signal level, with values $\sim$10$^{-1}$ mJy $\mathrm{beam^{-1}}$.  Any algorithm producing inaccuracies greater than this can cause the residual RMS to override the signal by order of magnitude, thereby obscuring it. This performance is consistently seen for both configurations.

\begin{figure*}
   \includegraphics[width=\columnwidth, height=8cm]{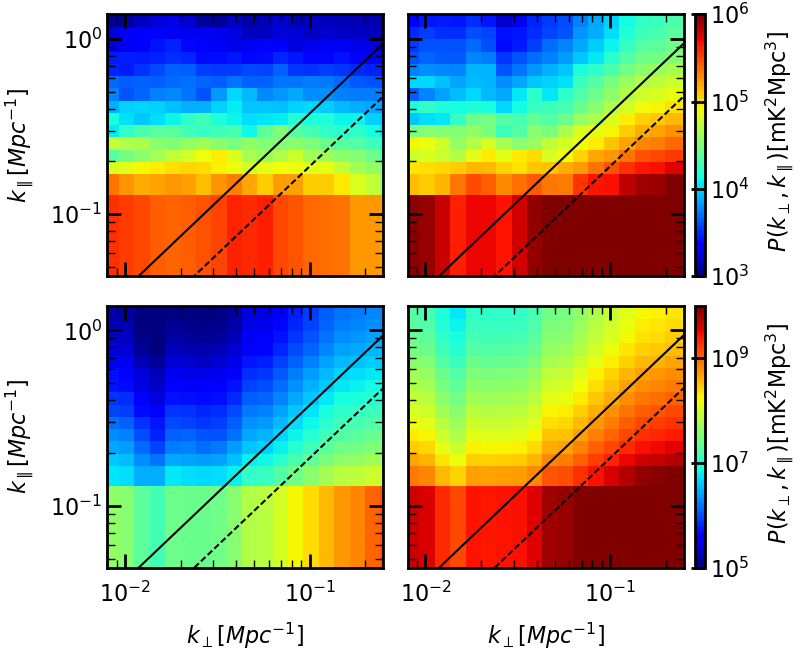}
    \includegraphics[width=\columnwidth, height=8cm]{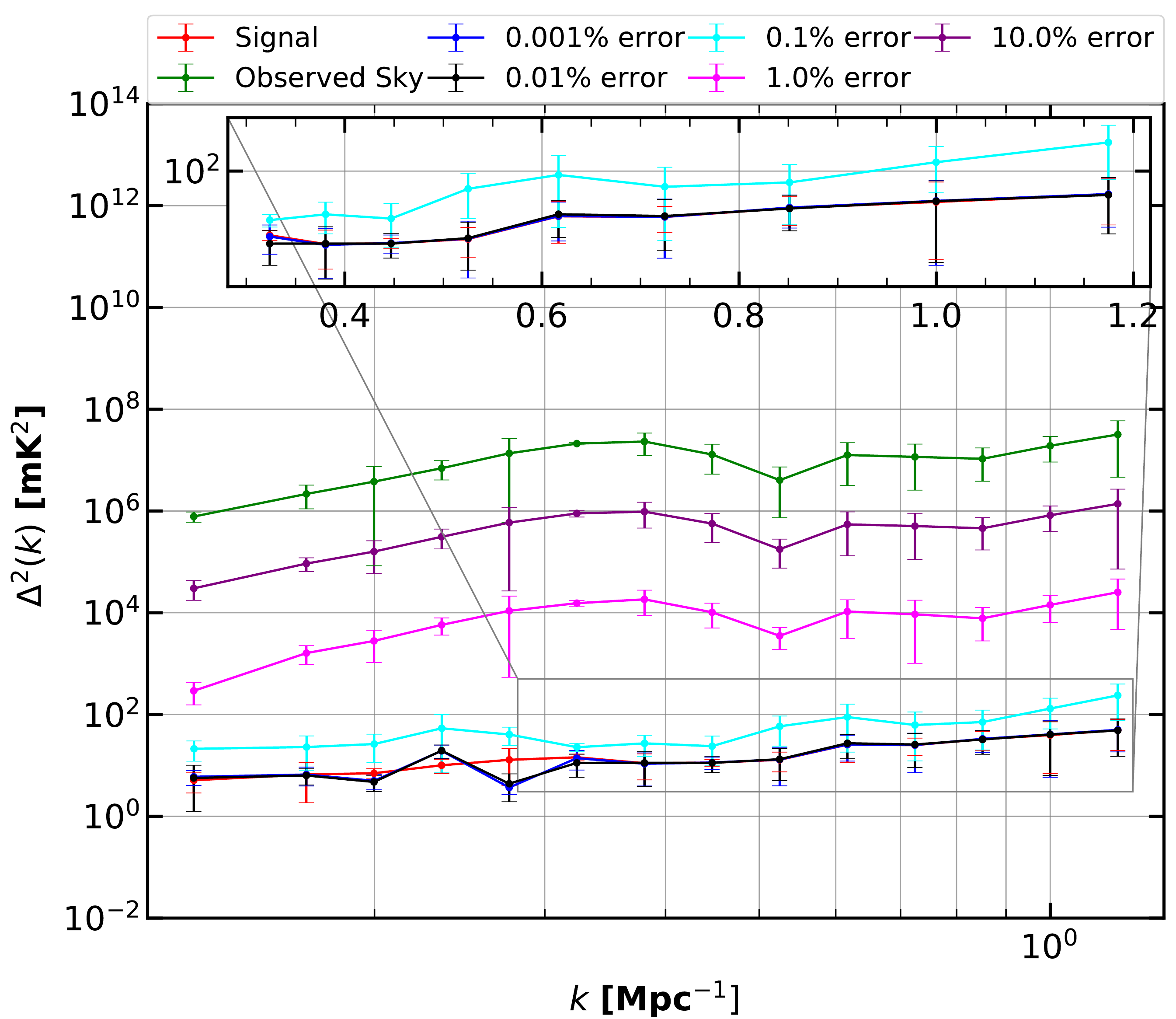}
  \caption{Residual power spectrum for gain errors with MWA Phase I. \textbf{(left)} 2D PS of the  residual visibilities for calibration errors for 0.001\% (top-left), 0.01\% (top-right), 0.1\% (bottom-left) \& 1.0 \% (bottom-right). The solid black line represents the horizon line, while the dashed black line is the FoV limit. \textbf{(right)} Spherically averaged PS with calibration residuals compared with the signal power and foreground power.The error bars are 3$\rm \sigma$ uncertainties for the k-bins including sample variance and thermal noise.} 
    \label{mwa_g}
 \end{figure*}
 
\begin{figure*}
   \includegraphics[width=\columnwidth, height=8cm]{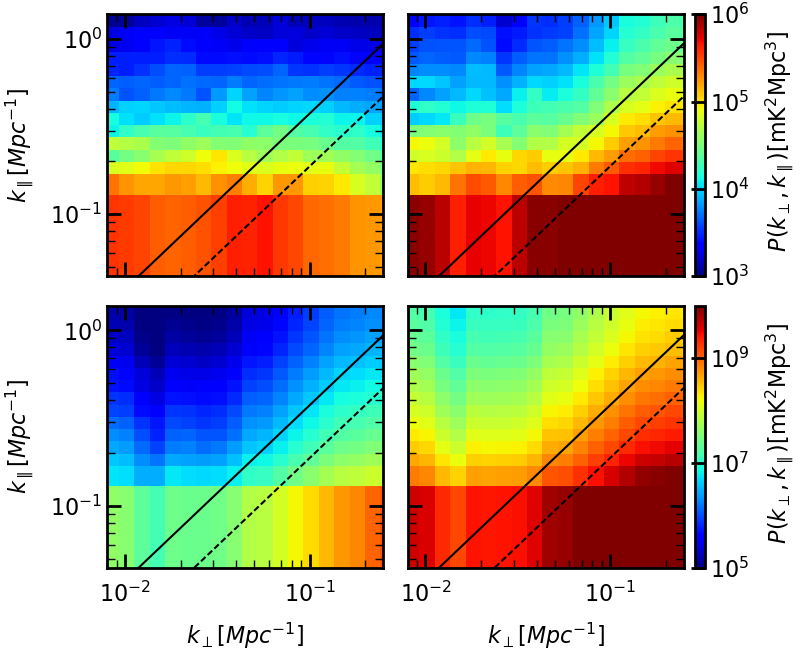}
    \includegraphics[width=\columnwidth, height=8cm]{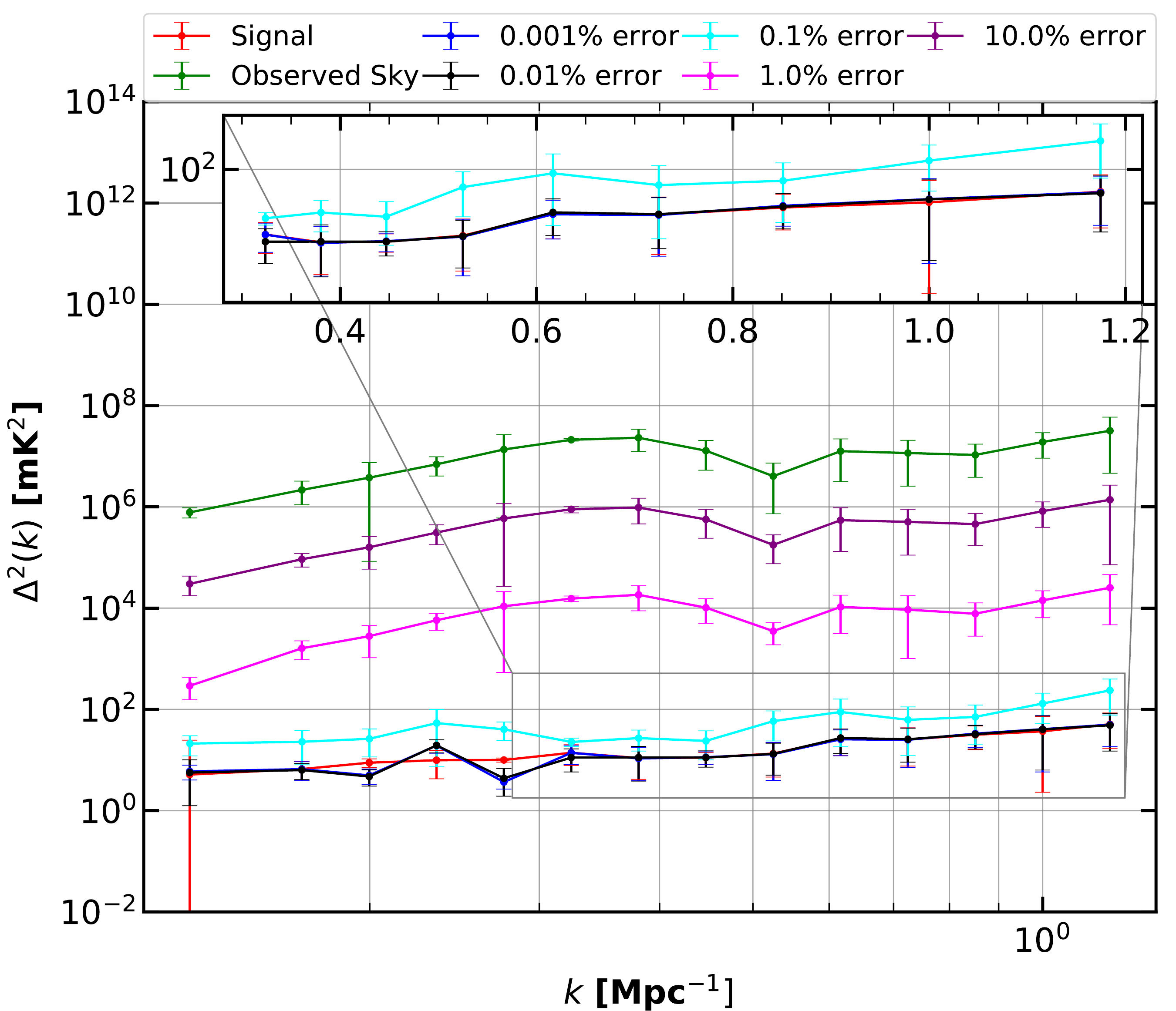}
  \caption{Residual power spectrum for gain errors with MWA-256. \textbf{(left)} 2D PS of the  residual visibilities for calibration errors for 0.001\% (top-left), 0.01\% (top-right), 0.1\% (bottom-left) \& 1.0 \% (bottom-right). The solid black line represents the horizon line, while the dashed black line is the FoV limit. \textbf{(right)} Spherically averaged PS with calibration residuals compared with the signal power and foreground power. The error bars are 3$\rm \sigma$ uncertainties for the k-bins including sample variance and thermal noise.} 
    \label{mwa1_g}
 \end{figure*}

The PS recovery performance for both MWA configurations is similar, which may be attributed to the fact that baselines within a maximum of 2 km of the core are used to determine the power spectrum (baselines of longer lengths are not of interest for EoR statistics).  Figures \ref{mwa_g} \& \ref{mwa1_g} show the PS recovered from the residual visibility of with calibration error using MWA-I \& MWA-256. For both cases, it is observed from the left panel that for the 2D PS, gain errors beyond 0.1\%, the residual foregrounds have power from the wedge leaking into the EoR window. Thus all the $k_{\perp}$ and $k_{\parallel}$ modes have higher power compared to the signal power, which implies the signal being obscured. The right hand panel for both Figures \ref{mwa_g} \& \ref{mwa1_g} show similar trends for the 1D power spectrum. For calibration inaccuracies $\leq$ 0.1\%, the residual power follows the signal power spectrum within error bars. This is also evident from the zoomed-in parts of both figures, where it can be seen that the residuals follow the signal power up to $\sim$0.01\%. 
Above 0.1\% , the residual foregrounds start to dominate over the signal exceeding the signal in orders of magnitude, thus obscuring it. 
  
\subsubsection{Position Error} 

The green and magenta curves in Figure \ref{pos_rms} show the residual RMS as a function of the angular displacement from the actual source position (top panel) and the relative percentage of error introduced (bottom) for MWA-I \& MWA-256, respectively. As seen from the green curves, in the case of MWA-I, an offset of 1\% (or $\sim$1\arcsec) or more with respect to the actual position results in the residual RMS level exceeding the signal level, thereby obscuring it. Similarly, for the magenta curve, i.e., MWA-256, the residuals lie within the signal level up to an offset of $\sim$ 0.1\arcsec, beyond which it exceeds the signal level and the thermal noise level for SKA1-Low, thereby rendering the signal undetectable.

\begin{figure*}
\includegraphics[width=\columnwidth, height=8cm]{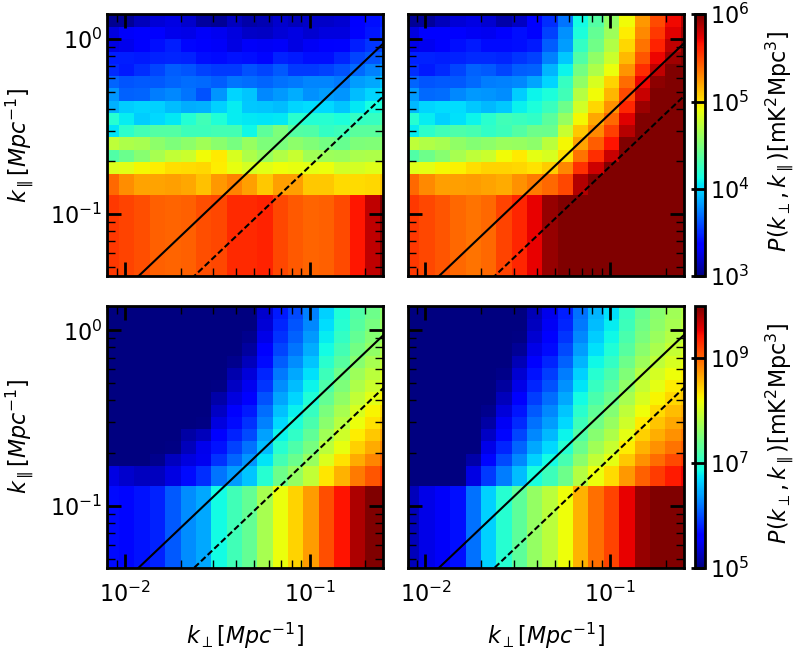}
\includegraphics[width=\columnwidth, height=8cm]{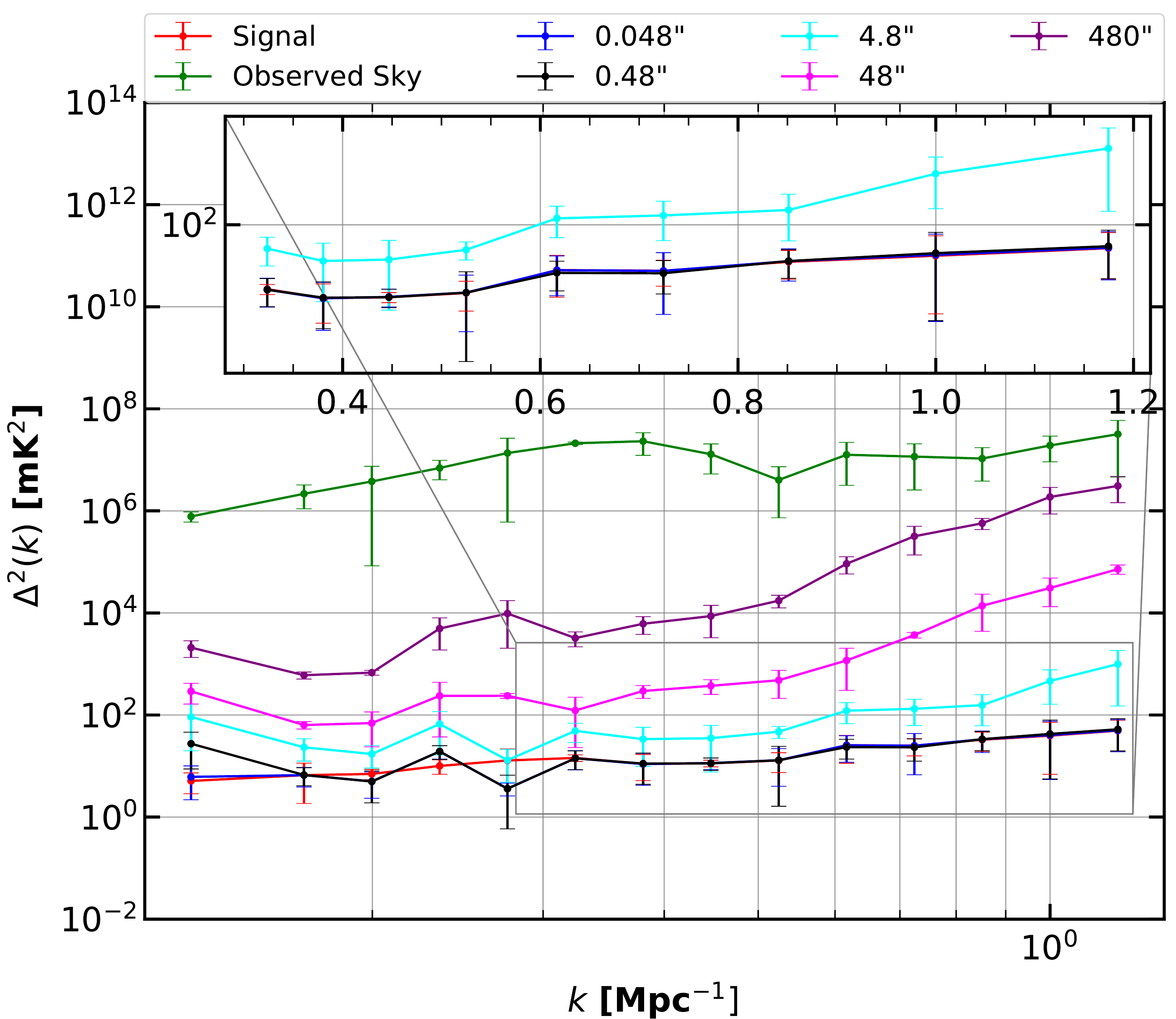}    
\caption{Residual power spectrum for position errors with MWA Phase I. \textbf{(left)} 2D PS of the  residual visibilities for position errors for 0.01\% (top-left), 0.1\% (top-right), 1.0\% (bottom-left) \& 10.0 \% (bottom-right). The solid black line represents the horizon line, while the dashed black line is the FoV limit. \textbf{(right)} Spherically averaged PS with position error residuals compared with the signal power and foreground power. The error bars are 3$\rm \sigma$ uncertainties for the k-bins including sample variance and thermal noise.}
\label{mwa1_p}
 \end{figure*}

\begin{figure*}
\includegraphics[width=\columnwidth, height=8cm]{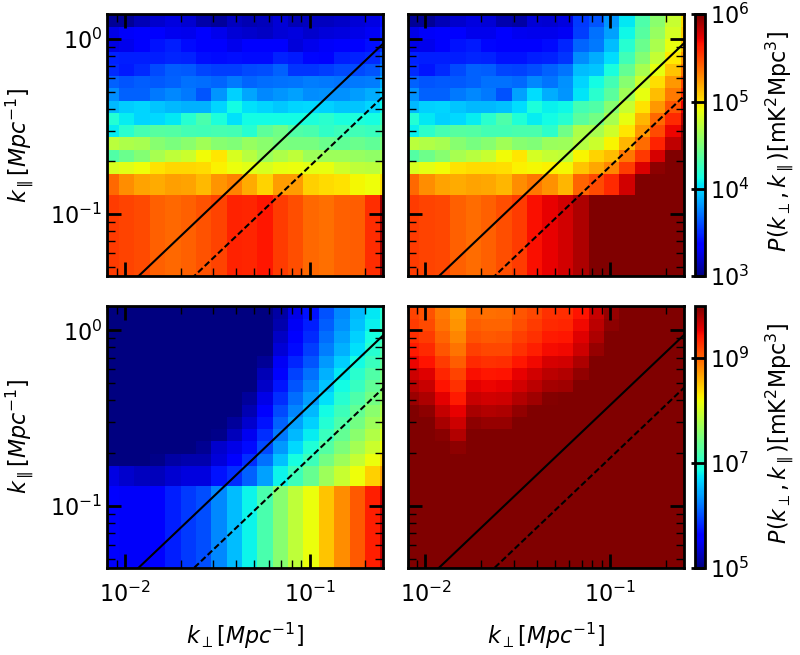}
\includegraphics[width=\columnwidth, height=8cm]{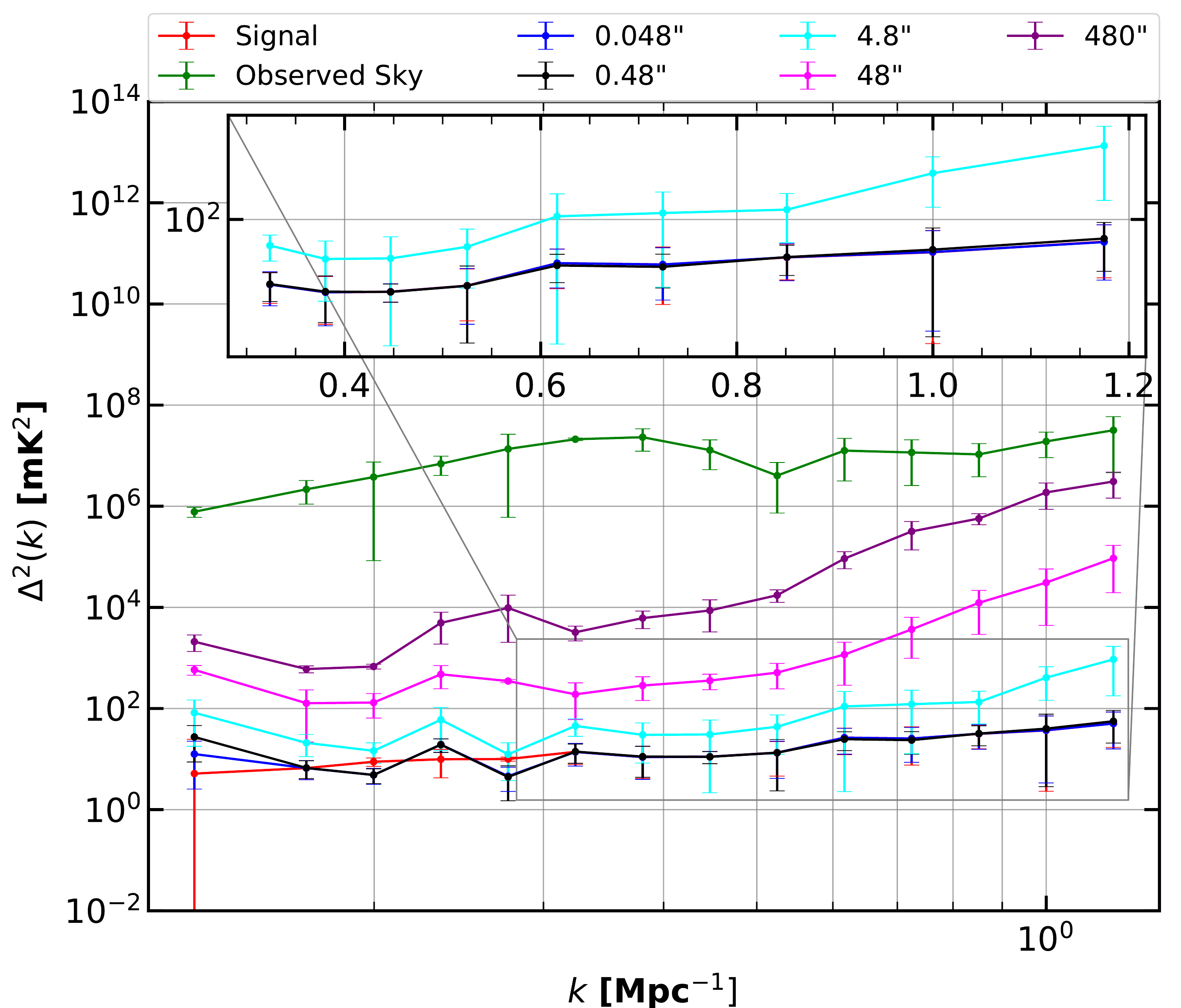}    
\caption{Residual power spectrum for position errors with MWA-256. \textbf{(left)} 2D PS of the  residual visibilities for position errors for 0.01\% (top-left), 0.1\% (top-right), 1.0\% (bottom-left) \& 10.0 \% (bottom-right). The solid black line represents the horizon line, while the dashed black line is the FoV limit. \textbf{(right)} Spherically averaged PS with position error residuals compared with the signal power and foreground power.The error bars are 3$\rm \sigma$ uncertainties for the k-bins including sample variance and thermal noise.}
\label{mwa_p}
 \end{figure*}

The residual PS for MWA configurations in the presence of position error are shown in Figures \ref{mwa1_p} \& \ref{mwa_p} - the left panel is the cylindrical PS, and the right panel is the spherical PS. As in the case of calibration errors, there is no significant difference between the two MWA configurations. From both the figures, it can be seen that in the entire $k_{\parallel}$-$k_{\perp}$ plane, the residual power has a higher magnitude than signal power at position error $\geq$1\% (i.e., displacement $\gtrapprox$0.1\arcmin). From the 1D PS (right panel of both figures), it is can likewise be observed that at 1\% error (cyan curve), the power spectrum of residual visibility starts deviating from the signal level (the residuals below this value follows the signal exactly). This implies that position accuracy better than 0.1\arcmin is required to detect the redshifted 21-cm signal from the EoR.

\subsection{Discussion}   

After comparing the performances of the arrays considered here in terms of both imaging and statistics (power spectrum), the following has been noted:

\begin{itemize}
    \item In the presence of residual calibration error, it is seen that for SKA1-Low,  inaccuracies of 0.001\%, there is sufficient DR in the image ($\sim$10$^{6}$) to detect the signal in the image domain. A careful inspection also reveals that for SKA1-Low, up to an error of $\sim$0.07\%, the residual RMS, though above the signal level, is below the daily 4-hour thermal noise level (orange line). The observations targeting EoR signal are expected to be systematics limited (rather than thermal noise limited).  Since the systematics are expected to be unrelated from day to day, reducing them becomes more critical than thermal noise. From this consideration, these simulations suggest that the image plane performance of SKA-1 Low is marginally better than the other arrays with identical observing strategy. However, it is worth mentioning that these simulations do not consider thermal noise. The noise for MWA is theoretically expected to be much higher compared to the other configurations under identical observation conditions. Nevertheless, detailed image plane analysis in the presence of noise will be pursued in future works.

    \item In the presence of position error, the accuracy of the order of $\gtrapprox$0.5\arcsec is required. At displacement 5\arcsec or more, the residual RMS lies above both the signal and 4-hour thermal noise level (orange line). This is consistently observed for all four array configurations. This level of astrometric precision is in principle achievable for observations using the SKA-1. Recent results from LOFAR and MWA show an astrometric accuracy of $\sim$1\arcsec (for example, the LoTSS survey between  120–168 MHz \citep{lotss} and the GLEAM survey between 70-231 MHz \citep{gleam1,gleam2}). Thus it is expected that the SKA-1, with improved sensitivity, should be able to produce sufficiently accurate sky models.

    \item For PS estimation, it is seen that residual calibration error causes the EoR window to get contaminated significantly for inaccuracies above 0.1\%. This trend is almost consistent in all the arrays considered and reflected in the spherically averaged PS. It is observed for the 1D PS that the power spectrum follows the signal reasonably well for errors below 0.1\%. After that, the residual amplitude in the $k_{\perp}$ - $k_{\parallel}$ plane starts exceeding the signal level. In the 1D PS, this manifests as deviation from the signal power at 0.1\% and above errors.
    
    \item From the PS estimates, it is seen that the residual calibration error causes significant contamination of the EoR window with calibration errors of $\sim$0.1\% and above. Below this value, the residual PS follow the "observed" signal power, and hence can be considered as the tolerance threshold for calibration errors.

    \item Residual PS in the presence of position error shows that for reasonably accurate astrometry, i.e., for a maximum position offset $\sim$0.5\arcsec, the residual power follows the signal power. Position offset of 5$\arcsec$ appears to be the threshold for tolerance, with some k modes lying within the signal uncertainty limits. Thus, it can be concluded that 5$\arcsec$ is the optimal tolerance for position error. Here, it should be reemphasized that the sky model used for data calibration is expected to be derived from higher-resolution observations. If they are not sufficiently accurate, the errors propagating into subsequent analyses would potentially confuse signal detection. 
    
    \item For the signal detection in PS domain, these simulations do not distinctly favour any of the four array configurations. However, it should be pointed out that these analyses are done under some very simplistic assumptions. In order to compare performances under even more realistic conditions, future works will explore various unavoidable systematics that can hamper observations targeting the redshifted 21-cm signal from CD/EoR.

\end{itemize}

\section{Influence of Array Specific Parameters on Calibration and Position Error}

This work considers the effects of residual calibration error and position error on the extraction of the 21-cm signal for four distinct array configurations. However, a detailed study of the influence of different array-specific parameters on the systematics considered here is beyond the scope of the present study. Nevertheless, this section briefly discusses a few parameters that vary across arrays depending on their location and specifications. 

With diameters of 40m, 14m, and 4m for SKA-1 Low, HERA, and MWA, the FoV is $\sim$ 3$^\circ$, 8$^\circ$ and 30$^\circ$ respectively at 142\,MHz. A larger FoV implies observation of a greater number of sources. Thus, an observation incorporating a large FoV results in more robust foreground statistics like source counts, clustering, etc. Such robust statistics are required for precision foreground modelling and removal. However, a larger FoV can result in the inclusion of bright sources in the telescope primary beam. Presence of extremely bright objects like the A-Team sources \citep{gleam1} in the southern sky and Cassiopeia A and Cygnus A in the northern sky hinder data calibration. Thus they need to be removed very carefully from the data \citep{mwa2}. A larger FoV may also result in inclusion nearby bright sources in the side lobes of the telescope beam. Their presence increase the already difficult problem of modelling and removal of foregrounds from EoR data. Additionally, any residual bright artifact generated by such sources can also obscure the 21-cm signal. Thus, the different FoV of the different arrays considered here provide both advantages and disadvantages.

Different FoV are also affected differently by ionospheric errors. The earth's ionosphere introduces phase corruptions to extragalactic signals at low frequencies ($\lesssim$1\,GHz). These effects are "directional", meaning they vary across different directions, causing an apparent shift of sources from their original positions by introducing additional phases. \citet{lonsdale2005} provided a description of the different regimes for performing ionospheric phase calibration, based on the observing FoV along with baseline lengths and irregularity size\footnote{\citet{lonsdale2005} designated the different regimes of calibration as regimes 1, 2, 3 and 4. The interested readers are directed to Figure 1 of the same.}. Variable ionospheric conditions can affect the arrays considered here differently \citep{lonsdale2005, Intema2009, wijnholds}. The compact array layout combined with the large beam size makes MWA fall on a calibration regime where there it is susceptible to having additional phases in the residuals due to ionosphere. This increases the probability of higher residual errors, even after self-calibration. The ionosphere-induced calibration inaccuracy in MWA EoR data was reported in \citet{jordan2017} and \citet{mwa2} used ionospheric metric to find usable data for providing upper limits on the EoR PS.  SKA-1 Low will be situated near the present MWA site and will face similar ionospheric conditions. However, each SKA station is projected to have a diameter of $\sim$40\,m, making its FoV smaller than the MWA at a similar frequency. Thus, it will fall in a different calibration regime of \citet{lonsdale2005}, where the smaller primary beam (i.e. FoV) is expected to introduce lesser ionospheric phase errors. For HERA, both the ionospheric conditions and FoV are different from SKA-1 Low and MWA, owing to its completely different location and dish diameter. However, a detailed study for HERA has not been done, and the analysis by \citet{2021arXiv210802263T} also does not consider any directional effects. Sensitive observations like those targeting the detection of the EoR need to understand the local ionosphere and calibrate its effects. But a detailed study into these effects is beyond the scope of the present work and is deferred till future works.

The baseline distribution as well minimum and maximum lengths of baselines also vary across arrays. The EoR signal is most sensitive at large angular scales and thus shorter baselines. The smallest $k_{\perp}$ mode accessible by an array is also determined by the minimum baseline length present. The arrays considered here have different baseline lengths and distributions. \citet{ewall2017} showed that the chromaticity increases with baseline lengths, causing longer baselines to have chromatic gain errors. So applying gain solutions to the shorter baselines mixes the contamination from the longer baselines to the shorter ones, which contaminates the EoR window. Thus, they suggest up-weighting the shorter baselines compared to the longer ones. \citet{ewall2017} also suggest that sky-based EoR experiments use short baselines to calibrate data used for PS estimation to prevent signal loss due to directional effects. Here, "short baselines" is a relative term depending on the array layout and the exact length (for example, $\sim$42\,m for SKA and $\sim$7\,m for MWA) should not matter. Longer baselines are essential for producing point source models for sky-based calibration. The recent  MWA Long Baseline Epoch of Reionisation Survey (LoBES) \citep{lobes} produced catalogues with better angular resolution than previous ones, which successfully removed foreground power from smaller angular scales. So it is evident that for successful detection of EoR PS, both long and short baselines are required for the best possible data processing.

The observing strategy also varies across arrays and is governed to a large extent by the actual physical properties of the dishes/tiles that make up the array. Most interferometers that are currently operational use tracking scans, i.e., tracks the phase center throughout the observation. Conversely, HERA is a drift scan array, meaning it does not track the sources but instead allows the sky to drift across its beam \citep{DeBoer2017}. The feasibility of doing drift scan observations for EoR PS estimation with the MWA has also been explored \citep{trott_2014, patwa2021}. The biggest advantage of drift scans is that the beam is not steered to track the source, thus providing better instrumental stability. Stable instruments are extremely important for sensitive observations targeting EoR signal. \citet{trott_2014} used MWA EoR experiment specifications to show that drift scans produce slightly lower uncertainties in the signal power and slightly higher SNR for the 1D PS compared to tracked observations. Simulations from \citet{trott_2014} also show that drift scans perform better in terms of reducing cosmic variance, thus providing better sensitivity at low k modes. But higher k modes for these observations are thermal noise dominated causing greater uncertainties due to the reduced coherence and subsequent increase in thermal noise.

A major challenge for drift scan observations is finding suitable calibrator sources to perform sky-based calibrations \citep{Kern_2020}. Thus, HERA has targeted to replace sky-based techniques with redundant calibration methods. MWA Phase 2 layout also has many redundant baselines for EoR observations. Still, both HERA and MWA use sky-based calibration at some point. The HERA team uses a final step of absolute calibration for breaking degeneracies of redundant calibration \citep{Dillon2020, 2021arXiv210802263T}. They build their models using bright sources from the GLEAM catalogue that coincides with the HERA track \citep{Kern_2020, 2021arXiv210802263T}. If such sources are not available, then calibration for drift scan observation constitutes a problem. It was also shown in \citet{Byrne2019, Kern_2020, Dillon2020} that the redundant calibration technique works only for very high redundancy and identical primary beam response. This was also seen by the HERA team while calibrating the observed data (see Section 3.2 of \citealt{2021arXiv210802263T}). Since it is impossible to have "perfect" redundancy, non-redundancies in a real array also lead to calibration errors. Thus, calibrating a drift scan instrument can be challenging due to the absence of "good" calibrator sources and imperfect redundancy. This may increase the calibration error and leak foregrounds into the EoR window. The observation and calibration strategy for SKA-1 Low is yet to be decided. Thus, it is worthwhile to simulate the merits and disadvantages of each method for SKA. But it should be reiterated that such detailed simulations are beyond the scope of the present work and are deferred till future works.

\section{Summary and Conclusion}
\label{conclusion}

Detection of the redshifted \hi\ 21-cm signal from the CD to EoR transitions is the most challenging undertaking for next-generation radio interferometers. The cosmological signal is weak, and prone to contamination by astrophysical foregrounds and instrumental systematics. The actual observations would require thousands of hours of data, which may be contaminated by improper calibration or position inaccuracy. Hence,it is extremely important to understand and quantify the effects of these systematic errors in order to be able to detect the cosmological signal. Hence, development of an end-to-end pipeline to check the effects of such systematic through simulations is essential. This work presents an end-to-end pipeline dealing with synthetic radio interferometric observations of the radio sky at low radio frequencies. The sky model includes the redshifted 21-cm signal and astrophysical foregrounds. Through various simulations the effects of these systematic errors in the extraction of the redshifted 21cm power spectrum has been shown. A comparison of results obtained for different array configurations- SKA1-Low, MWA-I, MWA-256, and HERA has also been demonstrated. The effect of the errors in the image plane detection of the cosmological signal has also been studied.

The simulations performed show that the optimal error in the case of calibration errors is 0.1\%, beyond which foreground domination in the residuals overrides the signal. This translates to a DR $\sim$10$^{5}$ or higher, which would require unprecedented precision in the algorithm employed. For the case of PS recovery, the signal is overridden by residual foregrounds at errors $>$0.1\%. Since the determination of the signal power spectrum is the primary aim for most interferometers, it can be said that the target precision for the calibration algorithm used should be of the order of 0.1\% or better. 

It is also found that position errors have a significant impact only if the displacement is $\geq$ 5\arcsec. PS derived from position error residual visibilities show that the residual power follows the signal power amplitude if the displacements are below $\sim$5\arcsec, valid for all the cases considered. Thus it is concluded that for the next generation interferometers, position errors alone should not be a major systematic limitation upto a reasonable level. However, it is also important to note that since position errors are not removable by calibration (being inherent in the sky model used in calibration), unless they are treated carefully, they may become a significant contaminant. 

The simulations performed here are done using a few simplistic assumptions for foregrounds and instruments. For such a scenario, they do not point to a preferred configuration among the four considered. The optimal tolerance for both calibration and position error are at a similar level for all four. However, it is seen that in presence of calibration error, the image plane performance of SKA and the PS performance of MWA-256 is marginally better compared to others. This points to the requirement of more detailed simulations with more complicated effects like primary beam chromaticity to determine the preferred array configuration. It should also be mentioned that for simplicity, the simulations are noise-free.  Hence, the limitations set by thermal noise have not been considered in the analysis. In our future works, we will incorporate thermal noise which will result in a more realistic case, to explore their impact on signal recovery.

%%%%%%%%%%%%%%%%%%%%%%%%%%%%%%%%%%%%%%%%%%%%%%
\section*{Acknowledgements}
AM thanks Indian Institute of Technology Indore for supporting this research with Teaching Assistantship. AM further acknowledges the NRAO staff \& the developers of OSKAR for promptly answering queries related to the use of simulation tool in CASA and the OSKAR software respectively. AM is thankful to Chandrashekhar Murmu and Mohd. Kamran for their help with the theoretical aspects of \hi\ signal, and also to Sumanjit Chakraborty for helpful discussions. AD would like to acknowledge the support from CSIR through EMR-II No. 03(1461)/19.
SM acknowledges financial support through the project titled ``Observing the Cosmic Dawn in Multicolour using Next Generation Telescopes'' funded by the Science and Engineering Research Board (SERB), Department of Science and Technology, Government of India through the Core Research Grant No. CRG/2021/004025.
The authors thank the anonymous reviewer and the scientific editor for helpful comments and suggestions that have helped to improve the quality of the work. 
%  AC would like to thank DST for INSPIRE fellowship. 
% This research made use of APLpy, an open-source plotting package for Python \citep{aplpy2012,aplpy2019} and Astropy,\footnote{\url{http://www.astropy.org}} a community-developed core Python package for Astronomy \citep{astropy:2013, astropy:2018}. 

\section*{Data Availability}
The simulated data for this study
will be shared upon reasonable request to the corresponding author.

{\bf $Software$ :}

This work relies on the Python programming language (\url{https://www.python.org/}). The packages used here are astropy (\url{https://www.astropy.org/}; \citealt{astropy:2013,astropy:2018}),numpy (\url{https://numpy.org/}), scipy (\url{https://www.scipy.org/}), matplotlib (\url{https://matplotlib.org/}). Simulations have been done using OSKAR (\url{https://github.com/OxfordSKA/OSKAR/releases}), Common Astronomy Software Applications CASA \url{https://casaguides.nrao.edu/index.php?title=Main_Page}.

%%%%%%%%%%%%%%%%%%%%%%%%%%%%%%%%%%%%%%%%%%%%%%%%%%


%\newpage
\bibliographystyle{mnras}
\bibliography{references} % if your bibtex file is called example.bib

%%%%%%%%%%%%%%%%%%%%%%%%%%%%%%%%%%%%%%%%%%%%%%%%%%

%%%%%%%%%%%%%%%%% APPENDICES %%%%%%%%%%%%%%%%%%%%%

\appendix
\section{}
 The power spectra with the 4 telescope configurations with EN1 as sky model for 0.1\% calibration and position errors. The trends observed with actual observed sky model is consistent with that obtained with a simulated sky model (i.e. T-RECS). The synthetic visibilities from both sky models show that the in the limit that only calibration or only position error is present, the optimum error should ideally be below 0.1\% for calibration errors and $\sim$1\arcsec for position. 
 \begin{figure}

\includegraphics[width=\columnwidth, height=8.0cm]{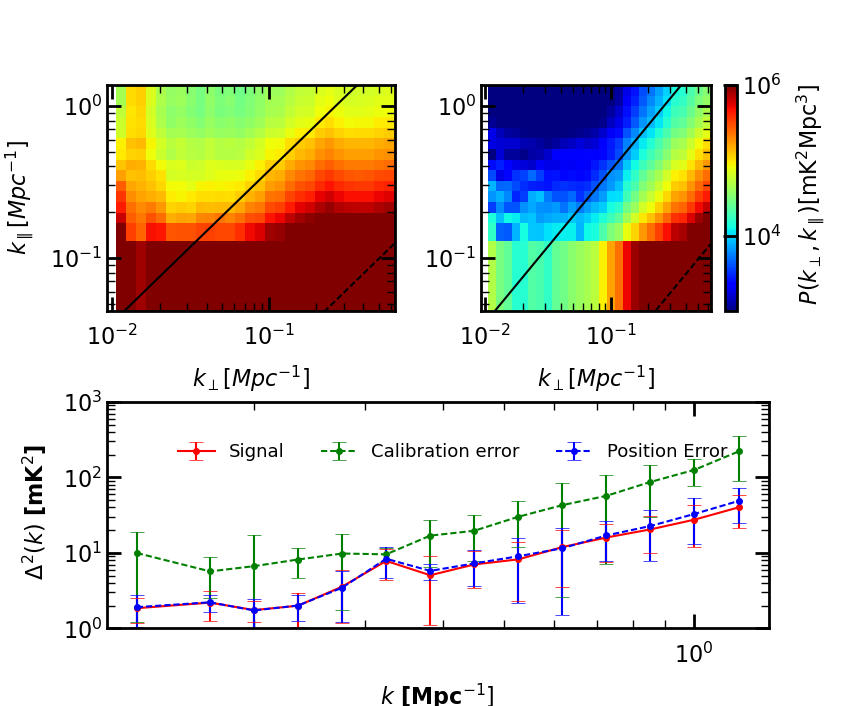}
\caption{Residual power spectrum with SKA-1 Low using EN1 as foreground model. \textbf{top-left} 2D PS with 0.1\% calibration error and \textbf{top-right} 2D PS with 0.1\% position error. \textbf{bottom} Spherical power spectrum for 0.1\% calibration error (green) and 0.1\% position error (blue).}
 \label{rep1}
  \end{figure}
  
  \begin{figure}

\includegraphics[width=\columnwidth, height=8.0cm]{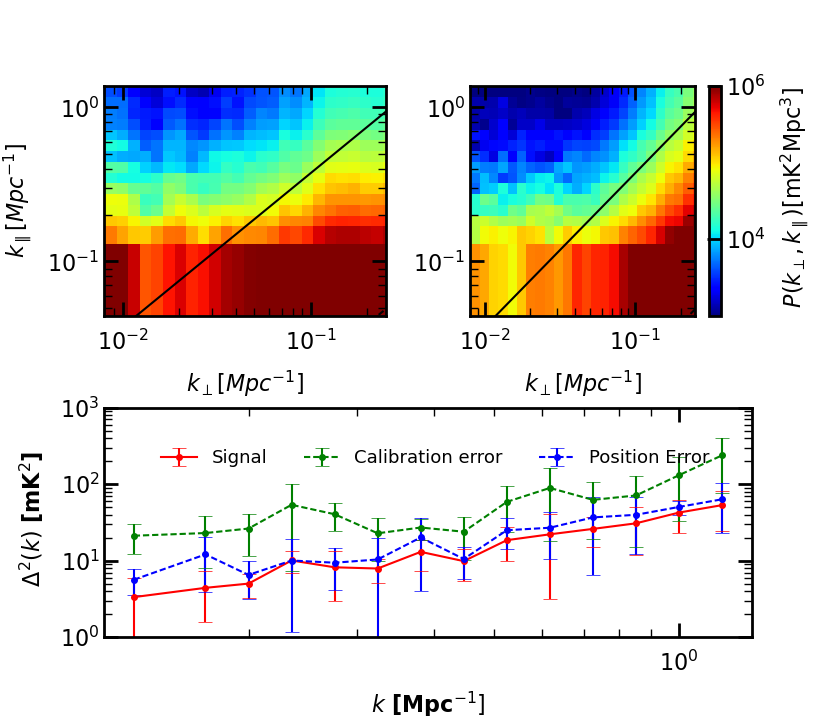}
\caption{Same as Figure \ref{rep1} using HERA.}
 \label{rep2}
  \end{figure}

\begin{figure}

\includegraphics[width=\columnwidth, height=8.0cm]{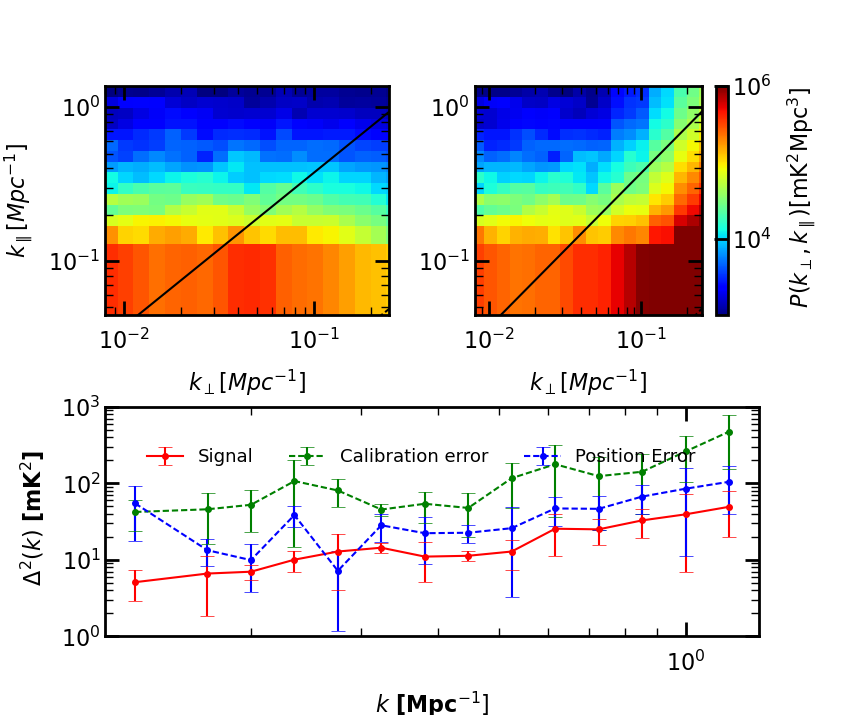}
\caption{Same as Figure \ref{rep1} using MWA-I.}
 \label{rep3}
  \end{figure}

\begin{figure}
 \includegraphics[width=\columnwidth, height=8.0cm]{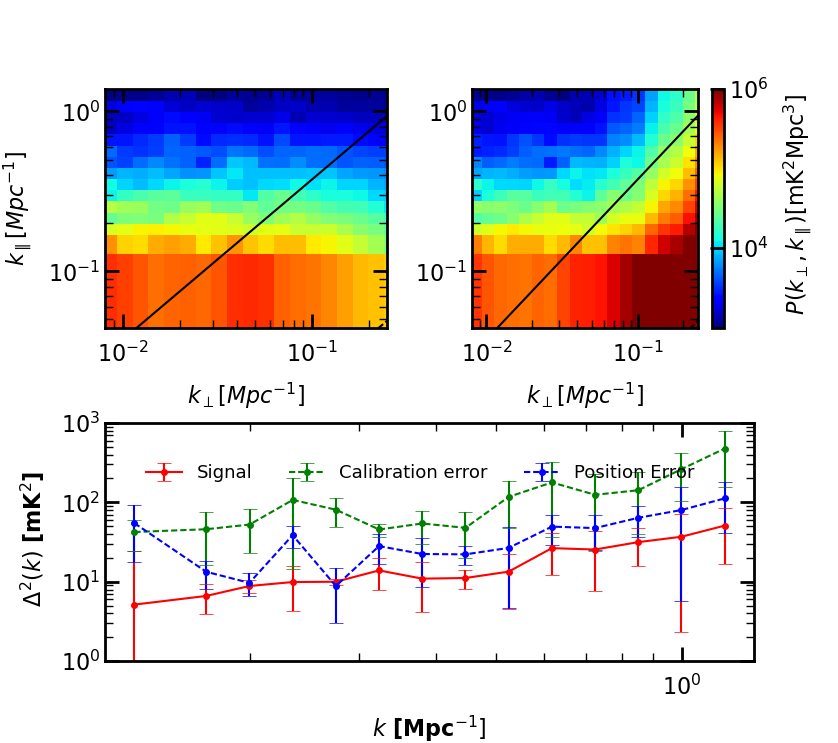}
\caption{Same as Figure \ref{rep1} using MWA-256.}
 \label{rep4}
  \end{figure}
\begin{figure}
 \includegraphics[width=\columnwidth, height=5.5cm]{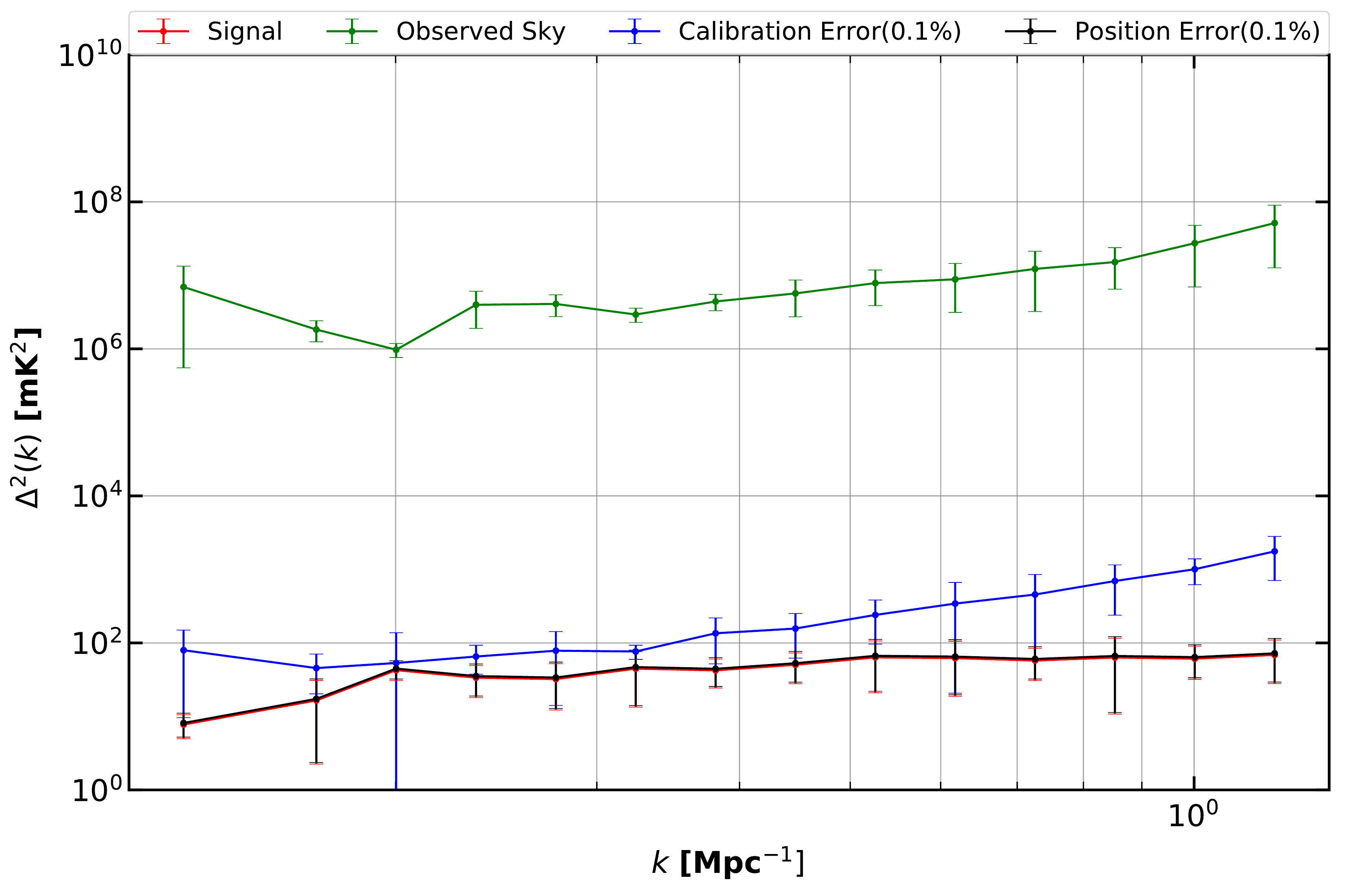}
\caption{Spherically averaged power spectrum as observed with SKA1-Low for signal only (red curve), calibration error residual (blue curve), position error residual (black curve) and the sky i.e. signal+foreground (green curve).}
 \label{rep5}
  \end{figure}

%%%%%%%%%%%%%%%%%
To demonstrate the capability of the pipeline to incorporate any signal model, Figure \ref{rep5} shows the spherically averaged power spectrum for observation using the Sem-Num model in four cases - observation of pure \hi\ only (red curve), the observed sky with signal and foreground (green curve), and residuals with 0.1\% calibration errors (blue curve) and 0.1\% position error (black curve). Comparison with the case for SKA1-Low (Figures \ref{ska_g} and \ref{ska_p}) shows that the results are consistent with those obtained for 21cmFAST signal model. This shows, once again the flexibility of the developed observational simulation pipeline to work with realistic scenarios.
 
% Don't change these lines
\bsp	% typesetting comment
\label{lastpage}
\end{document}